\documentstyle[emulateapj]{article}

\slugcomment{To appear in ApJ 2003 Oct 1}

\input psfig.sty

\newcommand{\beq}{\begin{equation}}
\newcommand{\eeq}{\end{equation}}

\newcommand{\cs}{c_{\rm s}}
\newcommand{\cmc}{~{\rm cm}^{-3}}
\newcommand{\cms}{~{\rm cm}^{-2}}
\newcommand{\Alf}{Alfv\'en\ }
\newcommand{\Msun}{~M_{\odot}}

\newcommand{\tc}{t_{\rm c}}

\newcommand{\kms}{~\rm km~s^{-1}}

\newcommand{\pc}{~\rm pc}

\newcommand{\K}{~\rm K}
\newcommand{\muG}{~\mu{\rm G}}

\begin{document}

\title{Nonlinear Hydromagnetic Wave Support of a Stratified
       Molecular Cloud}

\author{Takahiro Kudoh and Shantanu Basu}
\affil{Department of Physics and Astronomy, University of Western Ontario,
London, Ontario N6A 3K7, Canada}

\begin{abstract}

We perform numerical simulations of nonlinear MHD waves in a 
gravitationally stratified molecular cloud that is bounded by a
hot and tenuous external medium. We study the relation 
between the strength of the turbulence and various global properties 
of a molecular cloud, within a 1.5-dimensional approximation. Under
the influence of a driving source of Alfv\'enic disturbances,  
the cloud is lifted up by the pressure of MHD waves and reaches
a steady-state characterized by
oscillations about a new time-averaged equilibrium state. The nonlinear 
effect results in the generation of longitudinal motions and many shock 
waves; however, the wave kinetic energy 
remains predominantly in transverse, rather than longitudinal, motions. 
There is an approximate equipartition of energy between the 
transverse velocity and fluctuating magnetic field (as
predicted by small-amplitude theory) in the region of 
the stratified cloud which contains most of the mass; however, this 
relation breaks down in the outer regions, particularly near the 
cloud surface, where the motions have a standing-wave character.
This means that the Chandrasekhar-Fermi formula 
applied to molecular clouds must be significantly modified in such 
regions. Models of an ensemble of clouds show that, for various strengths 
of the input energy, the velocity dispersion in the cloud $\sigma 
\propto Z^{0.5}$, where $Z$ is a characteristic size of the cloud.
Furthermore, $\sigma$ is always comparable to the mean 
\Alf velocity of the cloud, consistent with observational results.

\end{abstract}

\keywords{ISM: clouds - ISM: magnetic fields - MHD - 
methods: numerical - turbulence - waves}

\section{Introduction}

Interstellar molecular clouds, the sites of 
current star formation in our Galaxy,
have long been known to yield supersonic line widths of
molecular spectral lines (e.g., see Zuckerman \& Palmer 1974).
Objects classified as molecular clouds span a large range of
mean radii ($R \sim 1 \pc - 100 \pc$), masses ($M \sim 10^2 \Msun -
10^6 \Msun$) and mean number density ($n \sim 10^1 \cmc - 10^3 \cmc$).
In fact, these quantities are correlated with the one-dimensional
velocity dispersion $\sigma$ through the well known 
line-width-size-density relations (e.g., Solomon et al. 1987):
\begin{eqnarray}
\sigma & = & 0.72 \, (R/{\rm pc})^{0.5} \; \kms, \label{eq:lwsize1}\\
n & = & 2.3 \times \, 10^3  \, (R/{\rm pc})^{-1} \; \cmc.
\label{eq:lwsize}
\end{eqnarray}
Thus, the velocity dispersion is typically supersonic since the 
sound speed $\cs$ is only $\approx 0.2 \kms$ for the typical molecular 
cloud temperature $T \approx 10 \K$ (e.g., Goldsmith \& Langer 1978).
 
The largest clouds, of mass $M \gtrsim 10^4 \Msun$, often
classified as Giant Molecular Clouds (GMC's) are in fact complexes
of smaller clouds, since the volume averaged density may be lower than
the excitation density of CO (e.g., Blitz \& Williams 1999), and also 
lower than allowed from thermal stability arguments (Falgarone \& Puget 1986). 
Hence, the basic building blocks of interstellar molecular clouds,
which contain most of mass of molecular material, are the 
dark (or dwarf) molecular clouds, which have
$R \sim 1 \pc - 10 \pc$, $M \sim 10^2 \Msun -
10^4 \Msun$ and $ n \sim 10^2 \cmc - 10^3 \cmc$. These clouds have
velocity dispersion $\sigma \sim 1 - 2 \kms$, and represent the class 
of objects that we are interested in modeling in this study.
We also note that observed smaller scale ($R \sim 0.1$ pc) dense cores 
are a separate object class which
are embedded within dark clouds and collectively contain only a small 
fraction of the total cloud mass.

While even dark clouds have masses that significantly exceed the 
thermal Jeans (1928) mass
\begin{eqnarray}
M_{\rm J} & = & \left( \frac{\pi k T}{mG} \right)^{3/2} \frac{1}{\rho^{1/2}} \nonumber \\ 
& = & 17 \left( \frac{T}{10 \K} \right)^{3/2}
\left( \frac{n}{10^3 \cmc} \right)^{-1/2}  \Msun
\end{eqnarray}
(where we have used $\rho = m n$, and $m = 2.33 m_{\rm H}$, in which 
$m_{\rm H}$ is the mass of a hydrogen atom), the
line-width-size-density relations do imply that molecular
clouds are individually in an {\it approximate} virial
balance between turbulent and gravitational energies.
In this paper, we equate the presence of nonthermal line widths
with the presence of a random superposition
of nonlinear (presumably hydromagnetic) waves, and refer loosely to the
latter as ``turbulence''.
Within each cloud, the turbulence
is expected to collectively exert a force (e.g., Chandrasekhar 1951)
which resists the inward pull of gravity.

The origin and persistence of turbulent motions in molecular clouds 
remain an active area of investigation. It was long understood 
(see Mestel 1965; Goldreich \& Kwan 1974) that supersonic 
hydrodynamic motions would 
decay rapidly through shocks, thereby creating a mystery of why the
turbulence was commonly observed throughout the lifetime of molecular clouds.
For GMC's, the lifetime is
estimated to be a few $\times \, 10^7$ yr
(e.g., Blitz \& Williams 1999), which is at least a few times longer than
the crossing time $\tc \equiv 2R/\sigma$ of the complexes.  The smaller 
dark molecular clouds 
may have even longer lifetimes $\sim 10^8$ yr (Shu, Adams. \& Lizano
1987).  Since slow and fast mode MHD waves are
also compressive and can be highly dissipative, it was suggested by 
Arons \& Max (1975) that the transverse \Alf mode might be a long-lived 
component, thereby preventing a rapid overall collapse 
of the clouds and explaining the observed persistence of the turbulence over 
their inferred lifetime. Mouschovias (1975) made the related suggestion
that the long-lived component may be due to standing waves,
i.e., normal mode oscillations, of magnetized clouds. Such global 
magneto-gravitational motions cannot be studied with a simple plane 
wave analysis in an infinite uniform medium.

The magnetohydrodynamic (MHD) wave picture has been strengthened by 
the detection of large-scale
magnetic fields within molecular clouds, through maps of polarized 
absorption and emission (e.g., Vrba, Strom, \& Strom 1976; Goodman et 
al. 1990; Schleuning 1998; Matthews \& Wilson 2002), and Zeeman measurements 
of the line-of-sight magnetic field strengths (Crutcher 1999 and references
therein). The latter imply that the magnetic energy (like the turbulent 
energy) is comparable to the gravitational
potential energy; equivalently, the mass-to-flux ratio
is close a critical value
\beq
\left( \frac{M}{\Phi} \right)_{\rm crit} = c_1 \: G^{-1/2}.
\eeq
In the above equation, the constant $c_1$ has been 
calculated to be in the range 0.13-0.17 based on detailed 
two-dimensional equilibrium
states (Mouschovias \& Spitzer 1976; Tomisaka, Ikeuchi, \& Nakamura 1988)
and is equal to $1/(2 \pi)$ in the case of a flattened one-dimensional
layer (Nakano \& Nakamura 1978). Collectively, the magnetic field measurements
allow the possibility that clouds are supported lateral to the large-scale
field $B$ by its associated Lorentz force, while these magnetic field lines
act as a carrier of Alfv\'enic disturbances with $\delta B \sim B$, 
explaining the observed spectral line widths and preventing the clouds
from assuming a very flattened configuration,

The dynamical effect of propagating MHD waves using analytic or
semi-analytic means has been studied by several authors.
Dewar (1970) developed a formalism for calculating the effect of
small amplitude hydromagnetic waves on a slowly varying background
medium, using the WKB approach. Assuming ideal MHD and no dissipation
of the waves, this leads to a steady-state relation between 
wave pressure $P_{\rm w}$ and gas density $\rho$ of the form
$P_{\rm w} \propto \rho^{1/2}$ (McKee \& Zweibel 1995).
Simplified calculations of the effect of small amplitude MHD waves on a 
molecular cloud have been performed by Fatuzzo \& Adams (1993) and Martin,
Heyvaerts, \& Priest (1997). Both satisfy the above scaling of $P_{\rm w}$
in the ideal MHD limit. In particular, the WKB model of Martin
et al. (1997) yields a steady-state density structure of an infinite
one-dimensional cloud supported by short-wavelength \Alf waves
in the ideal MHD limit, and also when accounting for damping of linear
waves by ion-neutral friction. In addition to ion-neutral friction,
which damps even linear waves (Kulsrud \& Pearce 1969), there are
several nonlinear effects which will work to enhance dissipation.
The second order effect of a gradient in the magnetic pressure 
$\nabla \; \delta B^2/8 \pi$ will in general lead to steepening of the 
waves followed by dissipation (Cohen \& Kulsrud 1974). 
Zweibel \& Josafatsson (1983) state the 
form of this dissipation rate, which can dominate the process of
ion-neutral friction for nonlinear and/or long wavelength modes.
There are also other known nonlinear avenues for wave dissipation,
such as the conversion of a parallel-propagating \Alf wave into an
acoustic wave and another \Alf wave traveling in the opposite direction
(Sagdeev \& Galeev 1969).

Recently, several studies have resulted in a solution of the 
full set of nonlinear equations of ideal MHD using finite difference
approximations. Gammie \& Ostriker (1996) performed a one-dimensional 
numerical simulation of MHD turbulence
in a periodic domain. This has been followed up by several multi-dimensional
simulations, also in a periodic domain
(e.g., Stone, Ostriker, \& Gammie 1998; Ostriker, Gammie \& Stone 1999; 
Mac Low et al. 1998; Mac Low 1999; Padoan \& Nordlund 1999; 
Ostriker, Stone, \& Gammie 2001).
These models impose prescribed velocity fluctuations, at the initial
time and sometimes also throughout the computed time, and investigate
the dissipation rate of turbulence, as well as various properties
of turbulent fluctuations.  One of the important results of these papers 
is that the decay time of the MHD turbulence 
is comparable to the crossing time over the driving scale of the turbulence.
The nonlinear coupling of the \Alf to the fast and slow mode MHD waves
is considered to be a significant 
source of the dissipation which reduces the lifetime of MHD turbulence 
relative to the ideal of pure Alfv\'enic turbulence.
We note that periodic models represent only a local region of a 
much larger molecular cloud, and maintain a fixed mean density. They 
cannot study global effects associated with the density stratification
in a cloud or the existence of a cloud boundary.

In this paper, we perform a different type of numerical simulation of 
MHD turbulence. We concentrate on one self-gravitating cloud and
study the effect of the turbulence on the mechanical structure of the cloud. 
This corresponds to an extension of the model studied by Martin et 
al. (1997) into a fully nonlinear counterpart.
As an initial condition, we use a hydrostatic 
equilibrium between thermal pressure and self-gravity in a cloud that
is bounded by an external high temperature medium.
We input turbulent energy into the system 
continuously and see how the mechanical equilibrium changes. 
Similar numerical simulations have been performed to study 
the propagation of Alfv\'en waves in the solar chromosphere and corona
(e.g., Hollweg, Jackson \& Galloway 1982; Mariska \& Hollweg 1985;
Hollweg 1992; Kudoh \& Shibata 1999; Saito, Kudoh, \& Shibata 2001). 
We extend this class of model to a self-gravitating cloud, and study 
the relation between the strength of the turbulence and various 
global properties of a molecular cloud. This is the first of a series 
of papers on global models of MHD wave support in molecular clouds. 
In this paper, we develop high-resolution one-dimensional models under 
the assumption of ideal MHD.

The paper is organized as follows.
The numerical model we used for the simulation is summarized
in \S\ 2. The results of the simulation are described in \S\ 3. 
We add some discussion of the results in \S\ 4 and
summarize the paper in \S\ 5.

\section{The Numerical Model}

\subsection{Schematic View of our Model}

Figure 1 shows a schematic picture of our model.
We consider a molecular cloud that is threaded by a
large-scale magnetic field,
and concentrate on a local region of the molecular cloud
enclosed by the rectangle in the figure.
We assume a driving force near the midplane of the cloud
and follow the dynamical evolution of the vertical structure 
of the cloud.

\bigskip
\bigskip
\psfig{file=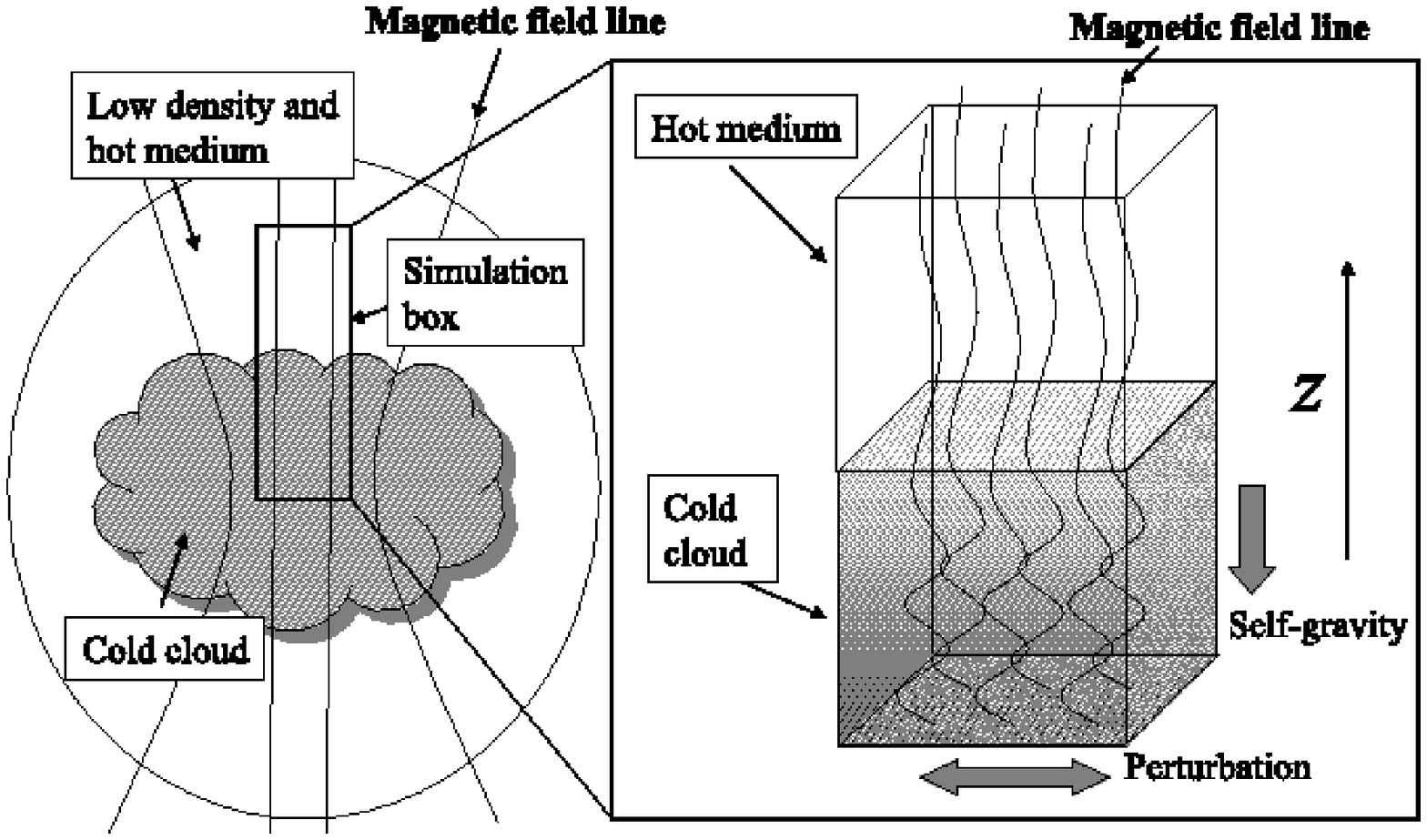,width=8.5cm}
\small
\bigskip\noindent {\sc Fig. ~1.~}\
Schematic picture of our model.
A molecular cloud that is threaded by a
large-scale magnetic field is considered.
A local region of the molecular cloud
enclosed by the rectangle in the figure can represent
the region we are modeling in this paper.
We input a driving force near the midplane of the cloud
and follow the dynamical evolution of the vertical structure
of the cloud.
\normalsize

\subsection{Assumptions}

For simplicity, we assume ideal MHD, 1.5-dimensions,
and isothermality for each Lagrangian fluid element.
The 1.5-dimensional approximation means that physical quantities depend
on only one coordinate, but we evolve nonzero components of vectors in
one additional direction. 
Isothermality for each Lagrangian fluid element means that the
temperature does not change in time for each fluid element
as it moves through Eulerian space, which is different from 
the assumption of a uniform time-independent temperature 
throughout the region.
These assumptions appear in \S\ 2.3 more concretely.

\subsection{Basic Equations}

We use local Cartesian coordinates $(x,y,z)$ on the molecular cloud, 
where we set $z$ to be the direction of the large-scale magnetic field. 
According to the symmetry of the 1.5-dimensional approximation, we set
\begin{equation}
\frac{\partial}{\partial x}=\frac{\partial}{\partial y}=0.
\end{equation}
The above symmetry and the divergence-free condition on the magnetic field
imply
\begin{equation}
B_{z} = {\rm constant},
\label{eq:Bz}
\end{equation}
where $B_z$ is the $z$-component of the magnetic field that 
threads the molecular cloud. Moreover, from the assumption of linear 
polarization of the waves, we can set
\begin{equation}
v_x=B_x=0
\end{equation}
without loss of generality, where $v_x$ and $B_x$ are the $x$-components 
of the velocity and magnetic field, respectively.

Therefore, the basic MHD equations we use in this paper are as follows:
mass conservation,
\begin{equation}
   {\partial \rho \over \partial t}
    + v_{z} {\partial \rho \over \partial z} 
    = -\rho {\partial v_{z} \over \partial z};
\label{eq:con}
\end{equation}
the $z$-component of the momentum equation,
\begin{equation}
   {\partial v_{z} \over \partial t}
    + v_{z} {\partial v_{z} \over \partial z}
    = - {1 \over \rho} {\partial P \over \partial z}
      - {1 \over 4 \pi \rho} B_{y} {\partial B_{y} \over \partial z} 
      + g_z;
\label{eq:vz}
\end{equation}
the $y$-component of the momentum equation,
\begin{equation}
   {\partial v_{y} \over \partial t}
    + v_{z} {\partial v_{y} \over \partial z}
    = {1 \over 4 \pi \rho} B_{z} {\partial B_{y} \over \partial z};
\label{eq:vy}
\end{equation}
the equation of energy,
\begin{equation}
   {\partial T \over \partial t}
    + v_{z} {\partial T \over \partial z}
    = -(\gamma-1) T \frac{\partial v_z}{\partial z};
\label{eq:eng}
\end{equation}
the $y$-component of the induction equation,
\begin{equation}
   {\partial B_{y} \over \partial t}
    = {\partial \over \partial z}( -v_{z}B_{y}+v_{y} B_{z});
\label{eq:By}
\end{equation}
the equation of state,
\begin{equation}
    P = \rho \frac{kT}{m} = c_s^2 \rho,
\label{eq:st}
\end{equation}
where 
\begin{equation}
c_s \equiv \sqrt{\frac{kT}{m}}
\end{equation}
is the isothermal sound speed;
and the Poisson equation,
\begin{equation}
    {\partial g_z \over \partial z}= -4\pi G\rho.
\label{eq:po}
\end{equation}
In these equations, $t$ is the time, $G$ is the gravitational constant, 
$k$ is Boltzmann's constant, $m$ is the mean molecular mass, 
$\gamma$ is the specific heat ratio,
$\rho$ is the density, $P$ is the pressure, $T$ is the temperature, 
$g_z$ is the $z$-component of the gravitational field, 
$v_z$ is the $z$-component of velocity, 
$v_y$ is the $y$-component of velocity, and 
$B_y$ is the $y$-component of the magnetic field. 

In equation (\ref{eq:eng}), we assume $\gamma=1$,
so that the energy equation becomes
\begin{equation}
   {\partial T \over \partial t}
    + v_{z} {\partial T \over \partial z}
    = 0,
\label{eq:tlag}
\end{equation}
which quantifies the assumption of 
isothermality for each Lagrangian fluid element.

\subsection{Initial Conditions}

As an initial condition, we assume hydrostatic equilibrium
of a self-gravitating one-dimensional cloud.
The hydrostatic equilibrium is calculated from the equations
\begin{equation}
\frac{1}{\rho}\frac{d P}{d z} = g_z,
\label{eq:hdsr}
\end{equation}
\begin{equation}
\frac{d g_z}{d z} = -4\pi G \rho,
\label{eq:hdsg}
\end{equation}
and
\begin{equation}
P = \rho \frac{kT}{m},
\label{eq:hdsp}
\end{equation}
subject to the boundary conditions
\begin{equation}
g_z(z=0)=0, 
\end{equation}
\begin{equation}
\rho(z=0)=\rho_0,
\end{equation}
and
\begin{equation}
P(z=0)= \rho_0 \frac{kT_0}{m},
\end{equation}
where $\rho_0$ and $T_0$ are the initial density and initial temperature 
at $z=0$, respectively.

In order to solve the above equations, we need to assume an 
initial temperature distribution.
If the temperature is uniform throughout the region,
we have the following analytic solution $\rho_{\rm S}$ found by Spitzer (1942); 
\begin{equation}
\rho_{\rm S} (z)=\rho_0 \, {\rm sech}^2 (z/H_0),
\end{equation}
where
\begin{equation}
H_0=\frac{c_{s0}}{\sqrt{2 \pi G\rho_0}}
\end{equation}
is the scale height, and 
\begin{equation}
c_{s0}=\sqrt{ \frac{k T_0}{m} }.
\end{equation}

However, an isothermal molecular cloud is usually surrounded by warm 
or hot material, such as neutral hydrogen or ionized gas.  
Therefore, we assume the initial temperature distribution 
to be
\begin{equation}
T(z)=T_0+\frac{1}{2}(T_c-T_0)\left[ 1+\tanh \left(\frac{|z|-z_c}{z_d} \right)\right],
\end{equation}
where we take $T_c=100T_0$, $z_c=3H_0$, and $z_d=0.2H_0$ throughout the paper.
This distribution shows that the temperature is uniform and equal to $T_0$
in the region of $0 \leq z < z_c=3H_0$ and smoothly increases to
another uniform value $T_c=100T_0$ at $z \simeq z_c=3H_0$.
By using this temperature distribution, we can solve the 
ordinary differential equations (\ref{eq:hdsr})-(\ref{eq:hdsp}) 
numerically. The numerical solution of these equations 
shows that the initial distribution of density is almost the same as 
Spitzer's solution in the region of $0 \leq z < z_c$ (see Fig. 2).

We also assume the following initial conditions;
\begin{equation}
v_z(z)=v_y(z)=0, 
\end{equation}
\begin{equation}
B_y(z)=0, 
\end{equation}
\begin{equation}
B_z(z)=B_{0}, 
\end{equation}
where $B_{0}$ is a constant. According to equation (\ref{eq:Bz}),
$B_z$ is spatially uniform and independent of time throughout the calculations. 

\subsection{Driving Force}

We introduce a perturbation into the initially hydrostatic cloud
by adding a driving force, $F(z,t)$,
into the $y$-component of the momentum equation (\ref{eq:vy}) 
as follows:
\begin{equation}
   \rho ({\partial v_{y} \over \partial t}
    + v_{z} {\partial v_{y} \over \partial z})
    = {1 \over 4 \pi} B_{z} {\partial B_{y} \over \partial z}
      + F(z,t),
\label{eq:driving}
\end{equation}
where
\begin{equation}
F(z,t) = \left\{
\begin{array}{ll}
\rho a_d (\frac{t}{10t_0}) \sin(2\pi \nu_0 t) \exp[-(\frac{z}{z_a})^2] & (t<10t_0) \\
\rho a_d \sin(2\pi \nu_0 t) \exp[-(\frac{z}{z_a})^2]  & (10t_0 \leq t \leq  40t_0) \\
0  & (t> 40 t_0), 
\end{array}
\right.
\label{eq:driving2}
\end{equation}
and 
\begin{equation}
t_0=H_0/c_{s0}.
\end{equation}
Furthermore, $a_d$ is the amplitude of the induced acceleration,
$\nu_0$ is the frequency of the driving force,
and $z_a$ represents the region in which we input the driving force.
The equations show that we input the sinusoidal driving force near 
the midplane of the cloud, and increase the maximum driving force 
linearly with time until $t=10t_0$, and maintain it to be 
constant during $10t_0 \leq t \leq  40t_0$. 
After $t=40t_0$, we terminate the driving force.

\subsection{Boundary Conditions}

We used a mirror symmetric boundary condition at $z=0$ and 
a free boundary at $z = z_{\rm out}$, the outer boundary of the calculation.
In order to remove the reflection of waves
at the outer boundary, we set $z_{\rm out}$ to be a large value, 
i.e., 
\begin{equation}
z_{\rm out} \simeq 2.6 \times 10^4 H_0,
\end{equation}
and use nonuniform grid spacing for large $z$ (see \S\ 2.8).

In addition to this, 
we set the gravitational field to be zero at large distances 
by introducing an artificial acceleration ($g_{-}$) into equations 
(\ref{eq:vz}) and
(\ref{eq:hdsr}), i.e.,
\begin{equation}
g_{-}(z) = -\frac{g_z(z)}{2}\left[ 1 + \tanh  \left(\frac{z-z_g}{5H_0}\right) \right],
\end{equation}
where we take $z_g =100H_0$ in this paper.
This force is zero until $z \simeq z_g=100H_0$, but 
it becomes essentially equal to $g_z$ with negative sign for $z >z_g=100H_0$
and compensates the gravity there.

The above boundary conditions show that we have effectively two 
outer boundaries.
The first one is $z_{\rm out}$ 
and the second one is $z_g$, which is the boundary for the gravitational field.
In order to remove the reflection of waves at the outer boundary,
it is useful to have a large $z_{\rm out}$. However, a very large
$z_{\rm out}$ introduces numerical problems because the density 
decreases exponentially at large $z$, according to the stratification of 
a self-gravitating cloud. A very low density leads to a very large
\Alf speed, making simulations inefficient by forcing very small time steps
for an explicit calculation.
By setting the net gravitational field to be zero at large distances,
we can avoid an extremely low density for large $z$ and 
therefore take a large $z_{\rm out}$ in order to 
remove the reflection of Alfv\'en waves at the outer boundary. 
We also note that for any slab of finite extent, as opposed to the 
infinite slab implied by our one-dimensional model, the net gravitational
field should indeed decrease at large distances, though not in the specific
manner prescribed here.

Due to the added artificial acceleration,
our effective numerical boundary is located at $z \simeq z_g=100H_0$, 
and we cannot trust the results beyond this region.
However, as we show later, all dynamical events 
we are interested in occur within $z < 50H_0$,
where most of the mass and energy are concentrated.
Hence, the effect of the artificial acceleration is 
negligible for the main results in this simulation.

\subsection{Numerical Parameters}

A natural set of fundamental units for this problem are
$c_{s0}$, $H_0$, and $\rho_0$. These yield a time unit
$t_0=H_0/c_{s0}$.
The initial magnetic field strength introduces one dimensionless
free parameter, i.e.
\begin{equation}
\beta_0 \equiv \frac{8\pi P_0}{B_0^2}=\frac{8\pi\rho_0 c_{s0}^2}{B_0^2},
\end{equation}
which is the initial ratio of gas to magnetic pressure at $z=0$.

In this cloud, $\beta_0$ is related to the mass-to-flux ratio.
For Spitzer's self-gravitating cloud, the mass-to-flux ratio normalized
to the critical value is
\begin{equation}
\mu_{\rm S} \equiv 2 \pi G^{1/2} \frac{\Sigma_{\rm S}}{B_0},
\end{equation}
where 
\begin{equation}
\Sigma_{\rm S}=
\int_{-\infty} ^\infty \rho_S \ dz =
2 \rho_0 H_0
\end{equation}
is the column density of the Spitzer's self-gravitating cloud.
Therefore,
\begin{equation}
\beta_0=\mu_{\rm S}^2.
\label{eq:bemu}
\end{equation}
The column density of the cloud we used in this paper is almost 
equal to that of Spitzer's cloud.
If we define the column density of the cloud, $\Sigma$, 
as the integral of density within  $-z_c <z< z_c$, then
\begin{equation}
\Sigma = \int_{-z_c} ^{z_c} \rho (t=0) dz \simeq 0.988 \, \Sigma_{\rm S}.
\end{equation}
This means that we can use the value of $\mu_{\rm S}$ an excellent 
approximation to the dimensionless mass-to-flux ratio of the model cloud.

Observations show that the dimensionless mass-to-flux ratios of molecular 
clouds are close to unity over a range of scales (Crutcher 1999; 
Shu et al. 1999).
Hence, we take $\beta_0=1$ in the models presented in this paper.

The driving force introduces three more important free parameters:
$\tilde{a}_d=a_d (H_0/c_{s0}^2)$, the dimensionless amplitude
of the acceleration due to driving, $\tilde{\nu}_0=\nu_0 t_0$,
the dimensionless frequency of driving, and
$\tilde{z}_a=z_a/H_0$, the dimensionless scale of the
driving region. For simplicity, we take $\tilde{\nu}_0=1$
and $\tilde{z}_a=0.1$ throughout this paper, and adjust 
the strength of driving by varying $\tilde{a}_d$ from $10$ to $50$.

Dimensional values of all quantities can be found through
a choice of $T_0$ and $\rho_0$, along with the values
of the dimensionless free parameters.
For example, if $T_0=10 \K$ and $n_0=\rho_0/m= 10^4 \cmc$,
then $c_{s0}=0.2 \kms$, $H_0=0.05 \pc$, 
$N_S=\Sigma_S/m=3 \times 10^{21} \cms$,
$t_0=2.5 \times 10^5$ yr, and $B_0=20 \muG$ if 
$\beta_0=1$.

\subsection{Numerical Technique}

In order to solve the equations numerically, we use the CIP method
(Yabe \& Aoki 1991) for equations (\ref{eq:con})-(\ref{eq:vy}) and
(\ref{eq:tlag}), and the MOCCT method (Stone \& Norman 1992) for 
equation (\ref{eq:By}).
The combination of the CIP and MOCCT methods is summarized in 
Kudoh, Matsumoto \& Shibata (1999).
The CIP method is a useful method to solve an advection equation 
such as equation (\ref{eq:tlag}) accurately, and is also
applicable to advection terms of equations (\ref{eq:con})-(\ref{eq:vy}).

Due to the mirror-symmetric boundary condition at $z=0$,
the Poisson equation (\ref{eq:po}) can be simply integrated from 
the midplane of the cloud;
\begin{equation}
g_z(z) = -4 \pi G \int_0^z \rho(z) dz. 
\label{eq:poi}
\end{equation}
We solve this equation by numerical integration.

In this simulation, we actually use variants of
the original CIP method.
We use a conservative-CIP method 
for the mass conservation equation (\ref{eq:con}), 
which was recently developed  by Xiao et al. (2002).
This scheme assures exact conservation of 
mass and less numerical oscillations. 
The original CIP method does not assure the
exact conservation of mass, so that a systematic deviation
during each time step can cause a problem for long time integrations;
this is especially a problem if the Poisson equation, which utilizes 
the mass distribution, is being solved simultaneously.
Hence, the conservative-CIP method significantly improves the accuracy
of our solution.
We also use the monotonic-CIP method (Xiao, Yabe \& Ito 1996) 
for the advection terms of equations (\ref{eq:vz}), (\ref{eq:vy}) and
(\ref{eq:tlag}), and use the CCUP method (Yabe \& Wang 1991) 
for the calculation of gas pressure, 
in order to get more numerically stable results. 
The recent developments of CIP related schemes are
summarized in Yabe, Xiao \& Utsumi (2001).

We used a uniform grid size, $\Delta z_i  = 0.02 H_0$, 
from $z=0$ to $z=80 H_0$, where $\Delta z_i$ is the grid size of
the $i$-th grid point. For $ 80 H_0 < z < 120 H_0$, we gradually 
increase the grid size as $\Delta z_{i+1} = Min\,(1.05 \Delta z_{i}, 0.2H_0)$,
where $Min$ is a function that picks the minimum from the two values;
hence, the maximum grid size in this region is $0.2H_0$.
For $120 H_0 < z < z_{\rm out}$, where the gravitational field is 
compensated by the artificial force, we further increase the grid size
according to $\Delta z_{i+1} = 1.05 \Delta z_{i}$.
In this paper, we used a total of 4310 grid points, 
most of which are concentrated in the region of 
the uniform grid, i.e., $z \leq 80H_0$.

\section{Results}

\subsection{Typical Result}

In this subsection, we show the results of $\tilde{a}_d=30$
as the typical case.

\subsubsection{Density Structure in the Cloud}

Figure 2 shows the density and temperature as a function of $z$. 
The dashed lines show the initial distributions and the solid lines
show the distribution at $t=30t_0$. 

\bigskip
\psfig{file=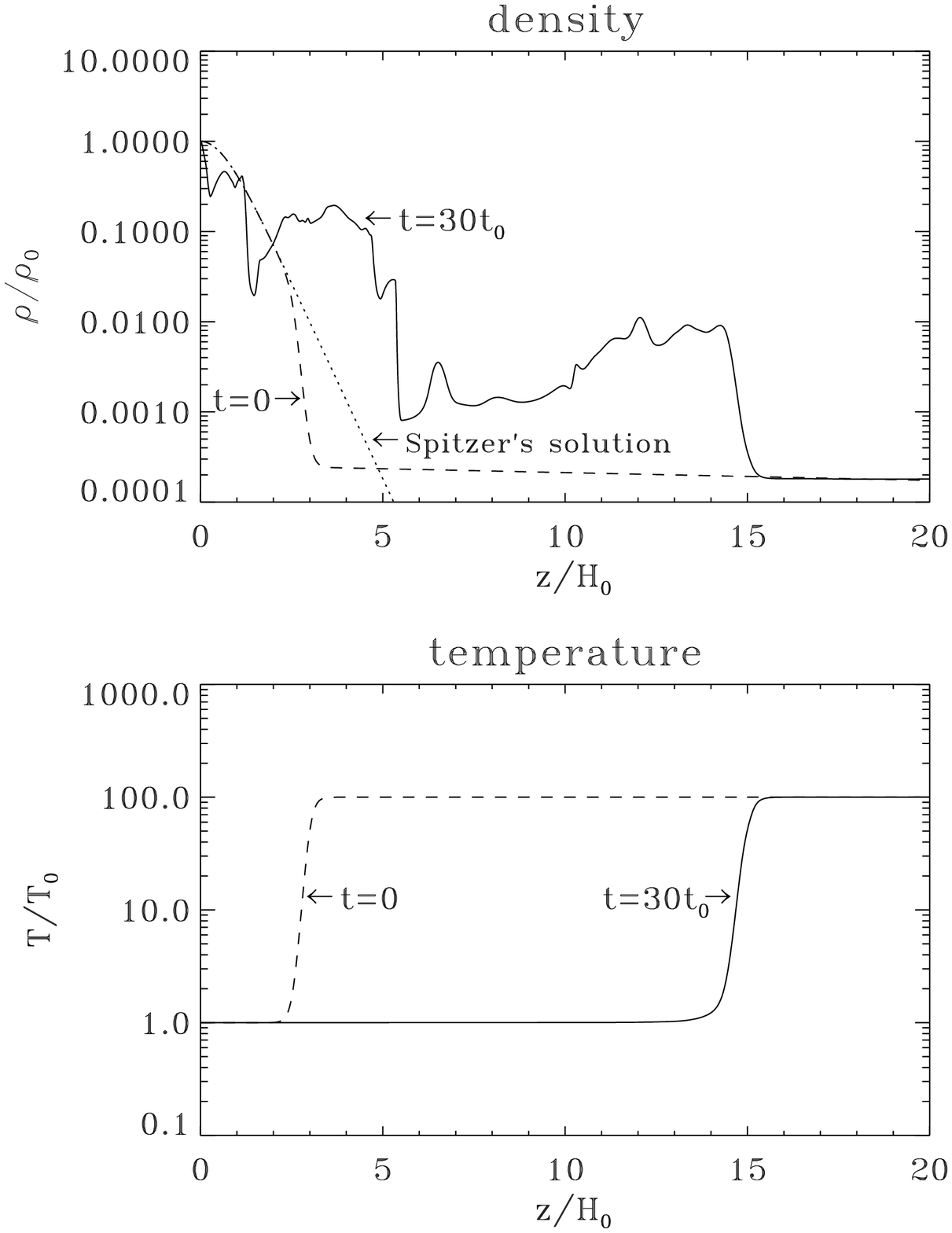,width=8.5cm}
\small
\smallskip\noindent {\sc Fig. ~2.~}\
Density (upper panel) and temperature (lower panel) as
a function of $z$. The dashed lines show the initial distributions
and the solid lines show the distribution at $t=30t_0$.
The dotted line in the density plot shows the distribution of
density in Spitzer's infinite self-gravitating equilibrium with a uniform
temperature $T_0$.
\normalsize
\bigskip

The dotted line in the density shows the distribution of 
Spitzer's self-gravitating cloud with uniform temperature of $T_0$.
The initial density 
deviates from Spitzer's solution around $z=z_c=3H_0$ where 
the initial temperature increases up to $100 T_0$. 
The density decreases rapidly around  $z=z_c=3H_0$ 
because of the pressure balance between low temperature cloud and
high temperature external medium. For $z>z_c=3H_0$, the scale height of 
the density is very large and the density decreases gradually.

The snapshot of density at $t=30t_0$ shows that 
the density has a complicated structure including many shock fronts.
This is caused by the driving force in equation (\ref{eq:driving}).
The driving force generates nonlinear Alfv\'en waves in the cloud
which produce a magnetic pressure gradient.
The magnetic pressure gradient and thermal pressure gradient usually push 
the cloud upward, but the self-gravity of the cloud always 
pulls it down. 
These up and down motions create the complicated structure
in the cloud. On the other hand, the temperature shows a smooth structure. 
This is due to isothermality for each Lagrangian fluid element.
Only the position of the temperature transition region 
changes in time.

Figure 3 shows the time evolution of the density.
The density plots at various times are stacked with time 
increasing upward in uniform increments of $0.2t_0$.
Because the driving force increases linearly with time up to 
$t=10t_0$, the density changes gradually at first. 
After $t=10t_0$, 
the density structure shows many shock waves 
propagating in the cloud, and significant upward and downward 
motions of the outer portion of the cloud, including the 
transition region.
After terminating the driving force at $t=40t_0$,
the shock waves are dissipated in the cloud and 
the transition region falls to around the initial position, 
although it is still oscillating.

\bigskip
\psfig{file=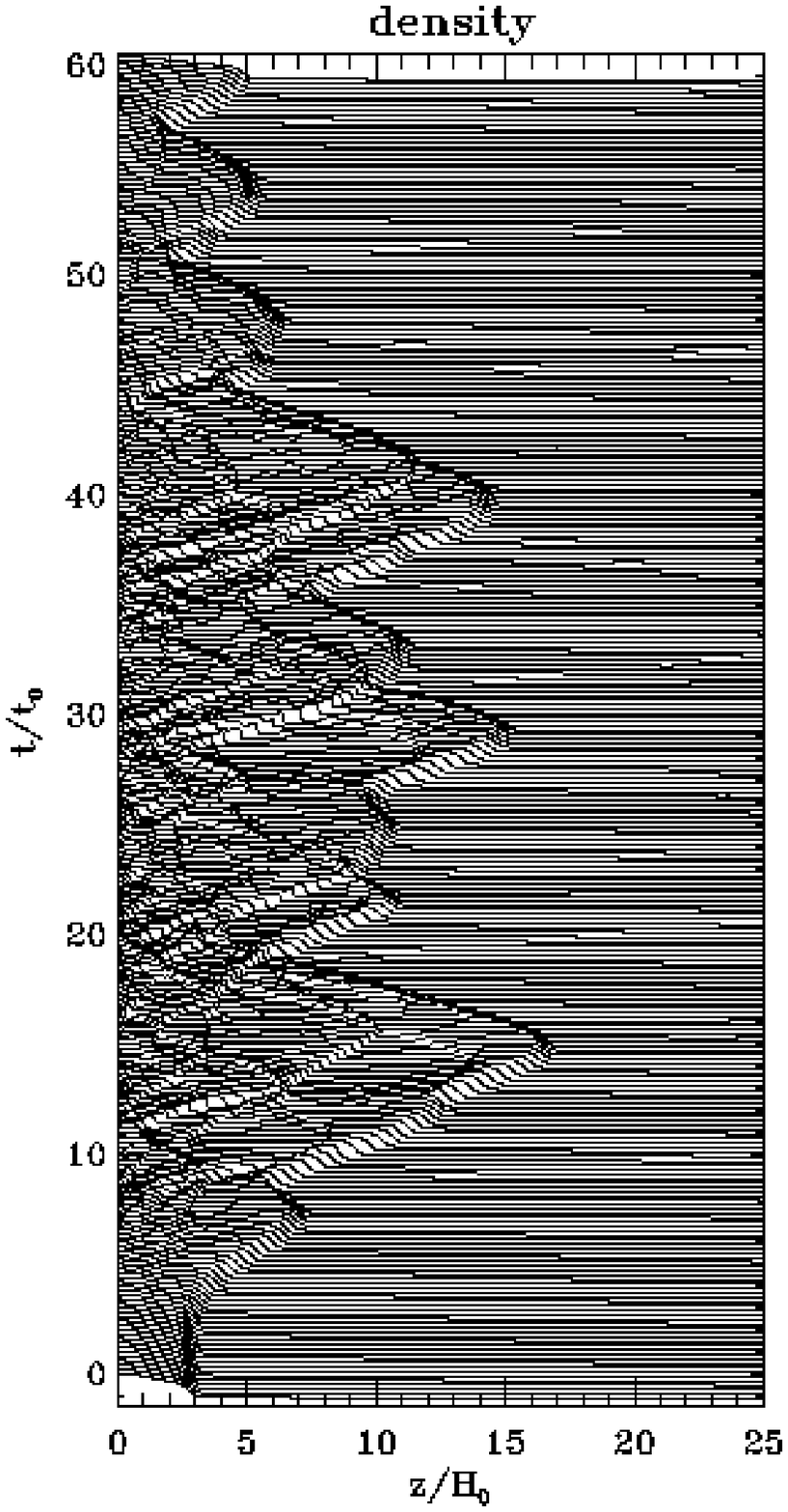,width=8.5cm}
\small
\smallskip\noindent {\sc Fig. ~3.~}\
Time evolution of the density.
The density plots at various times are stacked with time
increasing upward in uniform increments of $0.2t_0$.
\normalsize
\bigskip

Figure 4a shows the column density as a function of
the time averaged position of several Lagrangian fluid elements,
which are equally spaced at time $t=0$ with spacing 
$\Delta z =0.1H_0$ starting at $z=0.01H_0$.
The time average is calculated between $t=10t_0$ and $t=40t_0$,
while the driving force is input with constant amplitude.
The dashed line shows the initial distribution.
The Lagrangian fluid elements have constant enclosed 
column density as a function of time.
By using this property, we evaluate the location of 
Lagrangian fluid elements from the surface density.
The difference between the initial distribution and that 
of the time average shows that the cloud is lifted up. 
Figure 4b shows the time average of the density
for each element as a function of the time averaged 
position of each element.
These time averages were also calculated between $t=10t_0$ 
and $t=40t_0$.
In contrast to the snapshot in Figure 2, the time averaged
density structure shows a smooth distribution.
The dashed line shows the initial density distribution of the cloud.
The scale height of the time averaged 
distribution is about 3 times larger than the initial value.

\bigskip
\psfig{file=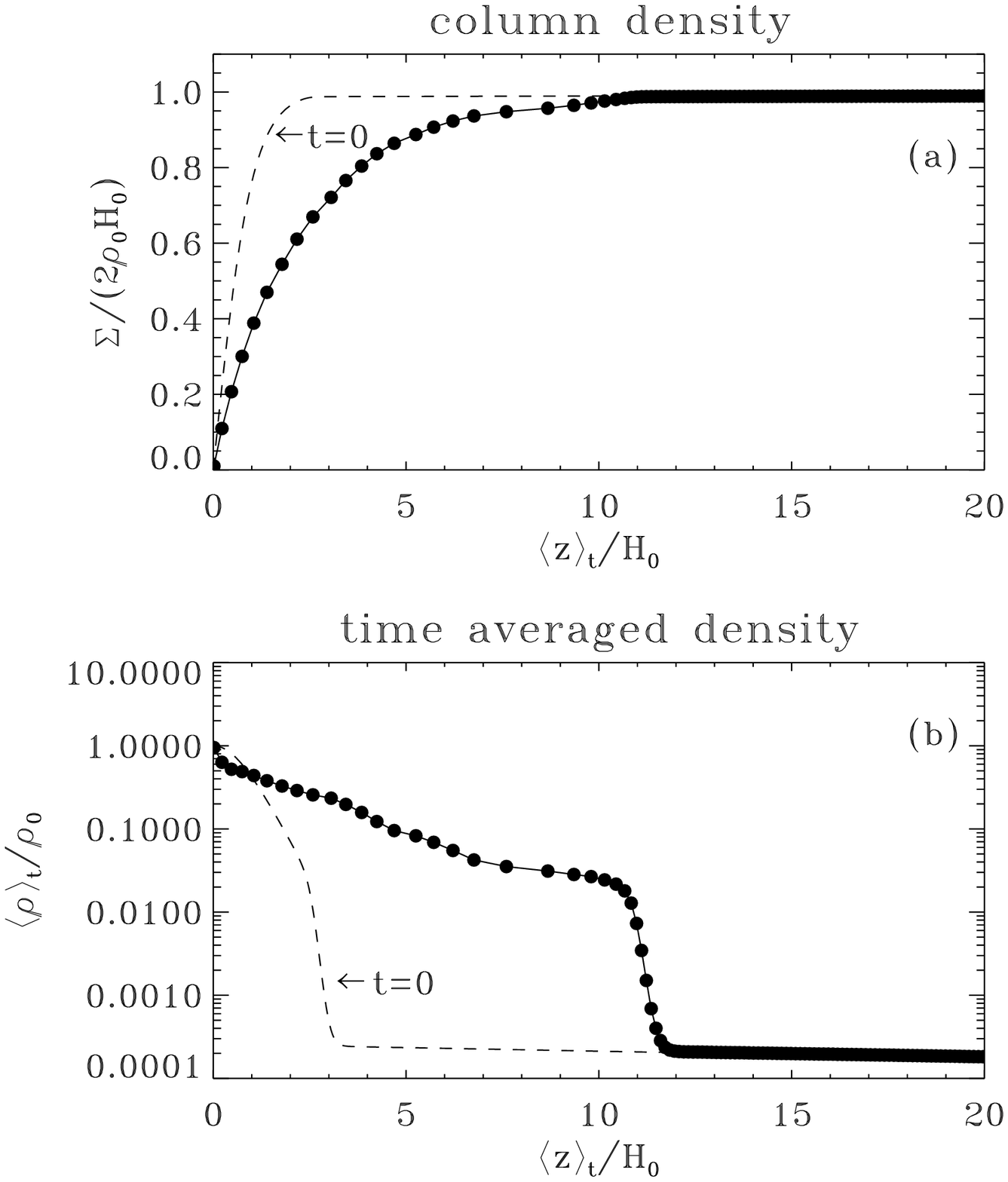,width=8.5cm}
\small
\smallskip\noindent {\sc Fig. ~4.~}\
(a) Column density as a function of the time averaged
position of each Lagrangian fluid element.
The dashed line shows the initial distribution.
The time average is calculated between $t=10t_0$ and $t=40t_0$.
The Lagrangian fluid elements have constant enclosed
column density as a function of time.
By using this property, we evaluate the location of
Lagrangian fluid elements from the surface density.
Each Lagrangian fluid element is equally spaced
at time $t=0$ with spacing $\Delta z =0.1H_0$ starting
at $z=0.01H_0$.
(b) Time average of the density for each fluid element
as a function of the time averaged position of the element.
The dashed line shows the initial distribution.
\normalsize
\bigskip

\subsubsection{Velocities in the Cloud}

Figure 5a shows the time evolution of $v_y$ at $z=0$.
Initially, it is oscillating about zero sinusoidally 
with the frequency of the driving force.
However, as time goes on, the oscillation shows sharp structures, 
and it becomes nonsymmetric around the mean.
The sharp structures are caused by nonlinear effects in the cloud
such as shock waves.
Moreover, the mean value shows a deviation from zero as time goes on.
This implies that 
the net $y$-momentum of the cloud is not zero and
the cloud has a mean motion in the $y$-direction 
as well as an oscillatory motion. 

\bigskip
\psfig{file=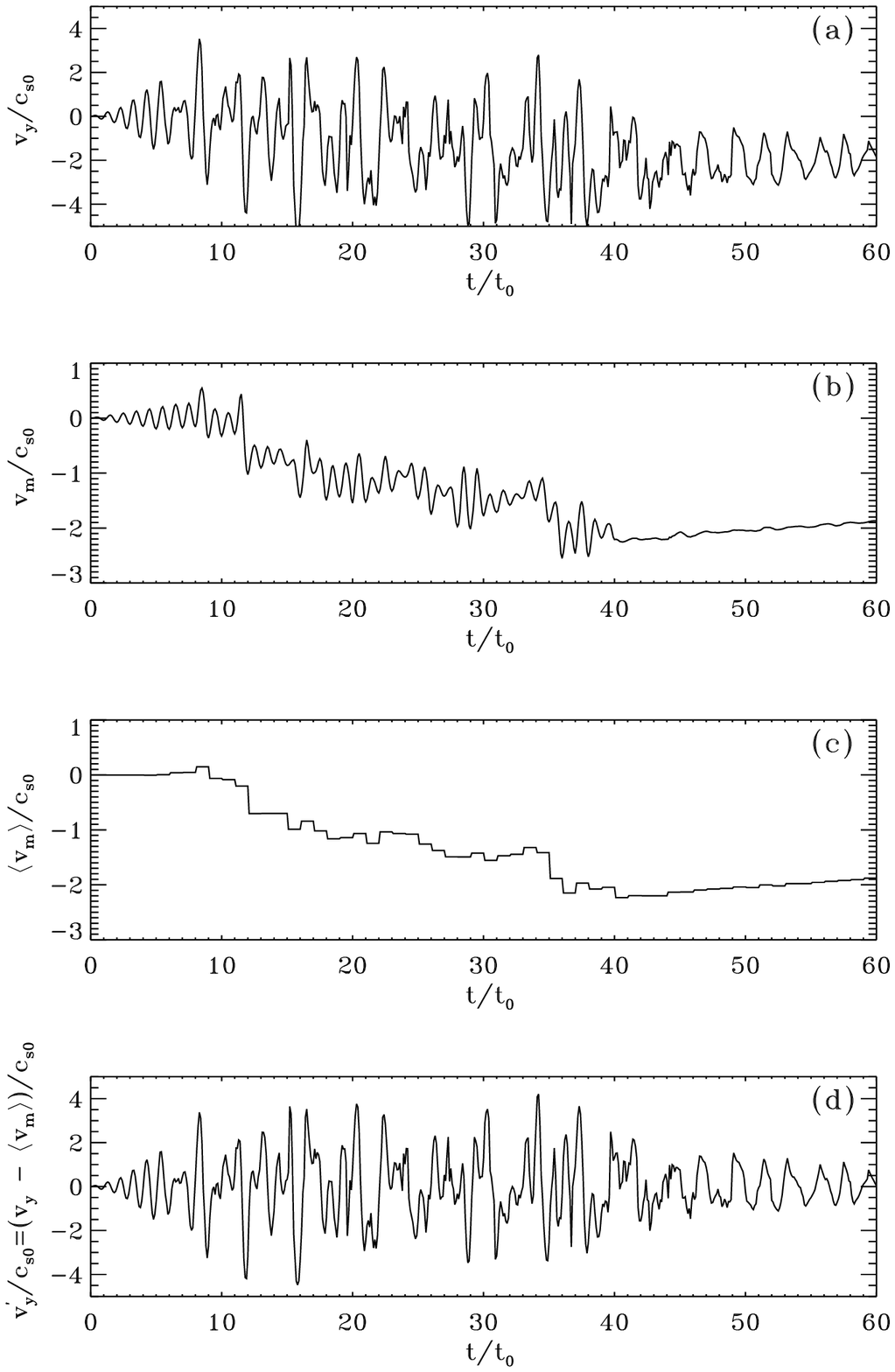,width=8.5cm}
\small
\smallskip\noindent {\sc Fig. ~5.~}\
Transverse velocities versus time.
(a) The $y$-component of the velocity at $z=0$ as a function of time.
(b) The mean $y$-component of velocity of the cloud as a function
of time.
(c) The time average of the mean $y$-component of velocity for each
cycle of the sinusoidal period of the driving force as a function
of time.
(d) The $y$-component of the velocity at $z=0$ minus
the time average of the mean $y$-component of velocity,
as a function of time.
\normalsize
\bigskip

This mean motion is ultimately caused by the driving force. 
Prior to $t=10t_0$, the driving force is not 
symmetric in time, because it is proportional to 
$t\sin(2\pi\nu_0t)$.
However, even if the driving force is symmetric in time,
the net momentum in the cloud never remains exactly zero
if we use a driving force like the one in equation (\ref{eq:driving2})
which is confined to a region near the midplane. The
net momentum of the cloud is generated by the nonlinear and 
nonsymmetric propagation of this disturbance into the stratified
cloud, the nonsymmetric restoring forces in the system, and 
nonsymmetric reflection at the cloud boundary.
As disturbances propagate, the restoring force due to
the magnetic field is not exactly symmetric in space and time. 
Also, some of the wave momentum is reflected at the 
boundary between the cold cloud and hot external medium, and part of it 
escapes from the cloud.
Hence, a net $y$-component of momentum can and does appear within
the cloud, creating a mean motion. 
We note that if we were modeling a two-dimensional cloud, i.e., 
the system had finite extent in the $y$-direction,
we could reduce the net drift with a $y$-distribution of driving 
force such that the net driving force is zero in the $y$-direction. 
Also, if the system was closed in the $z$-direction, as in a 
periodic boundary system, it is possible to add momentum in a regulated 
way so that the net transverse momentum remains exactly zero 
(e.g., Gammie \& Ostriker 1996).
However, the combined effects of our 1.5-dimensional approximation, cloud
stratification, and open cloud boundary make it difficult to 
regulate the net drift of the cloud in the $y$-direction.

Therefore, for our analysis, we divide $v_y$
into two parts. The first is the mean velocity which shows
the mean motion of the entire cloud.
The second is the oscillating component of the velocity.
We calculate the mean $y$-component of velocity of the cloud as
\begin{equation}
v_{\rm m}(t) = \frac{\int_0^{z_f(t)} \rho v_y dz}{\int_0^{z_f(t)} \rho dz}.
\end{equation}
In this equation, $z_f(t)$ is the full mass position of the cloud, 
which is defined by
\begin{equation}
\int_0^{z_f(t)} \rho dz = 0.998 \frac{\Sigma_{\rm S}}{2},
\end{equation}
and $z_f(t)$ corresponds to the position of the Lagrangian fluid element
which is initially located at $\simeq z_c$, the
initial position of the transition region of the temperature.
The time evolution of $v_m$ is shown in Figure 5b.

Figure 5b shows that the mean velocity is still oscillating 
while the driving force is input ($t=0-40t_0$).
In order to remove the oscillation,
we take time average of $v_m$ for each
cycle of the sinusoidal period ($t_0$) of the driving force
and define $\langle v_m\rangle$ in the following manner.
For example, $\langle v_m\rangle$ between $nt_0$ and $(n+1)t_0$
is calculated as 
\begin{equation}
\langle v_m\rangle(t=nt_0-(n+1)t_0) = \frac{1}{t_0} 
\int_{nt_0}^{(n+1)t_0} v_{\rm mean}(t^\prime) dt^\prime,
\end{equation}
where $n$ is an integer. This calculation is done from $n=0$ to $n=59$.
Figure 5c shows $\langle v_m\rangle$ as a function of time. 
According to the definition,
$\langle v_m\rangle$ has the same value between $nt_0$ and $(n+1)t_0$.

Finally, we define the oscillating component of 
the $y$-velocity as
\begin{equation}
v^\prime_y=v_y-\langle v_m\rangle.
\end{equation}
Figure 5d shows the time evolution of $v^\prime_y$.
In contrast to Figure 5a, it is oscillating around
$v^\prime_y=0$.

Figure 6 shows the time average of various velocities for
each Lagrangian fluid element plotted versus
the time average of the position of
each element.
Open circles show the time averaged root mean square of $v_z$, 
filled circles show that of $v^\prime_y$,
and the dashed line shows the time average of the sound speed.
The solid line shows the root mean square of the time average 
of the effective one-dimensional velocity dispersion
\begin{equation}
\sigma=\sqrt{\onehalf [v_z^2+(v^\prime_y)^2] + c_s^2}
\end{equation}
for each fluid element.
The time average is taken between $t=10t_0$ and $t=40t_0$.
This figure shows that the $y$-component of the velocity is 
the dominant component in the cloud. 
The averaged $y$-component of the velocity increases as 
a function of $\langle z\rangle_t$, except for near the midplane. 
This means that the largest velocity dispersion occurs in the
low density region. This is a similar tendency to that of 
linear Alfv\'en waves. If we assume the WKB approximation
and no wave dissipation, the energy flux of the waves is constant, i.e.,
\begin{equation}
\rho \, (v^\prime_y)^2 \, V_A = {\rm constant},
\end{equation}
where $V_A$ is the Alfv\'en velocity of the background magnetic field,
i.e., 
\begin{equation}
V_A = \frac{B_z}{(4\pi\rho)^{1/2}}.
\end{equation}
This leads to
\begin{equation}
v^\prime_y \propto \rho^{-1/4}.
\label{eq:vyWKB}
\end{equation}
Although this relation is not exactly applicable to 
our nonlinear result, our result shows a similar tendency.
We discuss this further in the next section.

\bigskip
\psfig{file=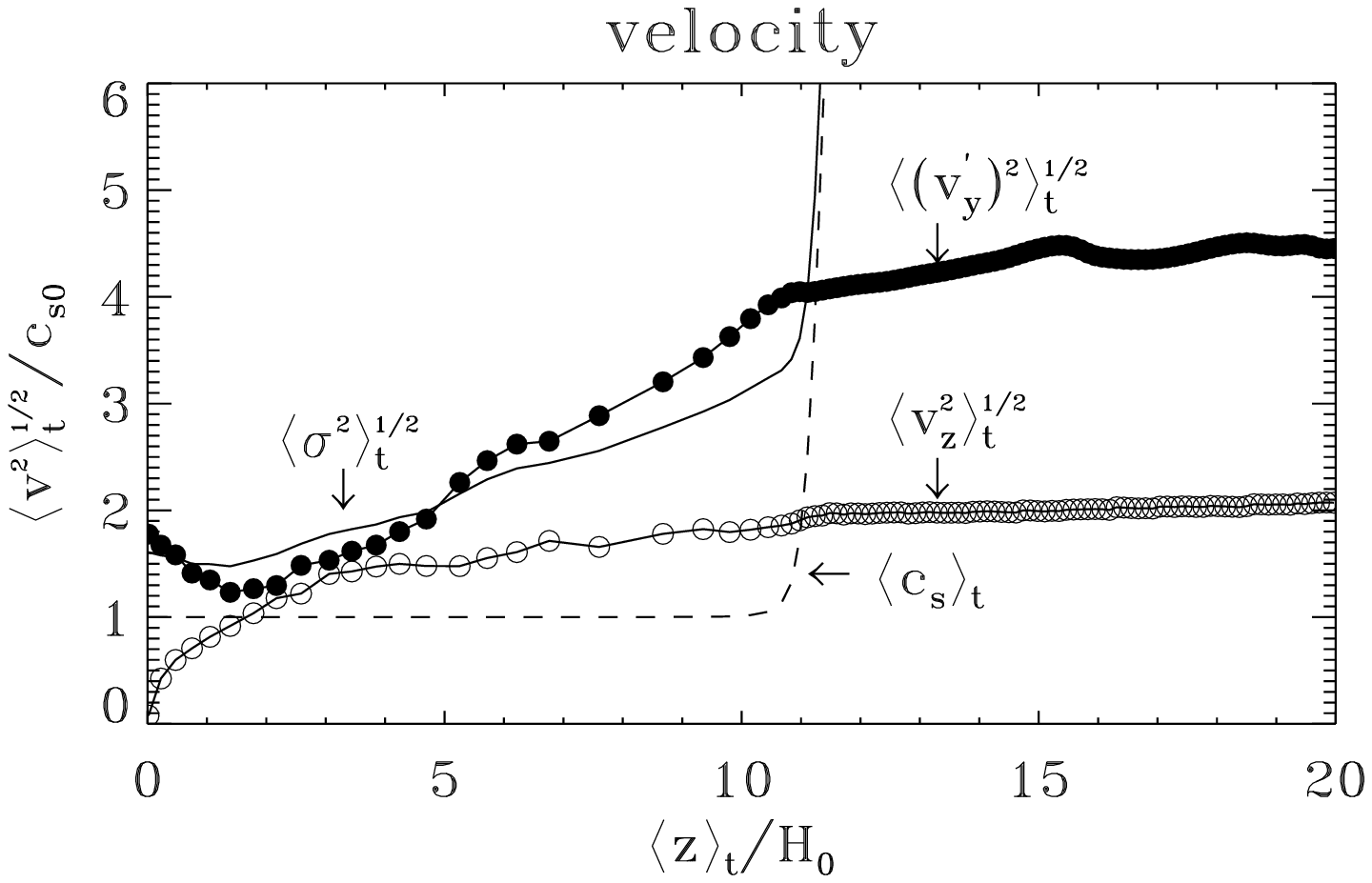,width=8.5cm}
\small
\smallskip\noindent {\sc Fig. ~6.~}\
Time averaged root mean square of various velocities for
each Lagrangian fluid element versus
the time average of the position of each element.
Open circles show the time averaged root mean square of $v_z$,
filled circles show that of $v^\prime_y$,
and the dashed line shows the time average of the sound speed.
The solid line shows the root mean square of the time average
of the effective one-dimensional velocity dispersion.
\normalsize

\subsubsection{Pressures in the Cloud}

Figure 7a shows the time averages of thermal pressure, magnetic pressure,
and dynamic pressures for each Lagrangian fluid element as 
a function of the time average of the position of each element.
The thick dashed line shows the thermal pressure, the thick solid line 
shows the magnetic pressure of the $y$-component of the magnetic field, i.e.,
$\frac{B_y^2}{8\pi}$, the dash-dotted line shows the dynamic pressure 
for the $y$-component of velocity, i.e., $\frac{1}{2} \rho (v^\prime_y)^2$,
and the dotted line shows the dynamic pressure of the $z$-component
of velocity, i.e., $\frac{1}{2} \rho v_z^2$. 
The thin straight line shows the magnetic pressure of background 
magnetic field, $\frac{B_0^2}{8\pi}$, and the thin dashed line shows the
initial thermal pressure.

\bigskip
\psfig{file=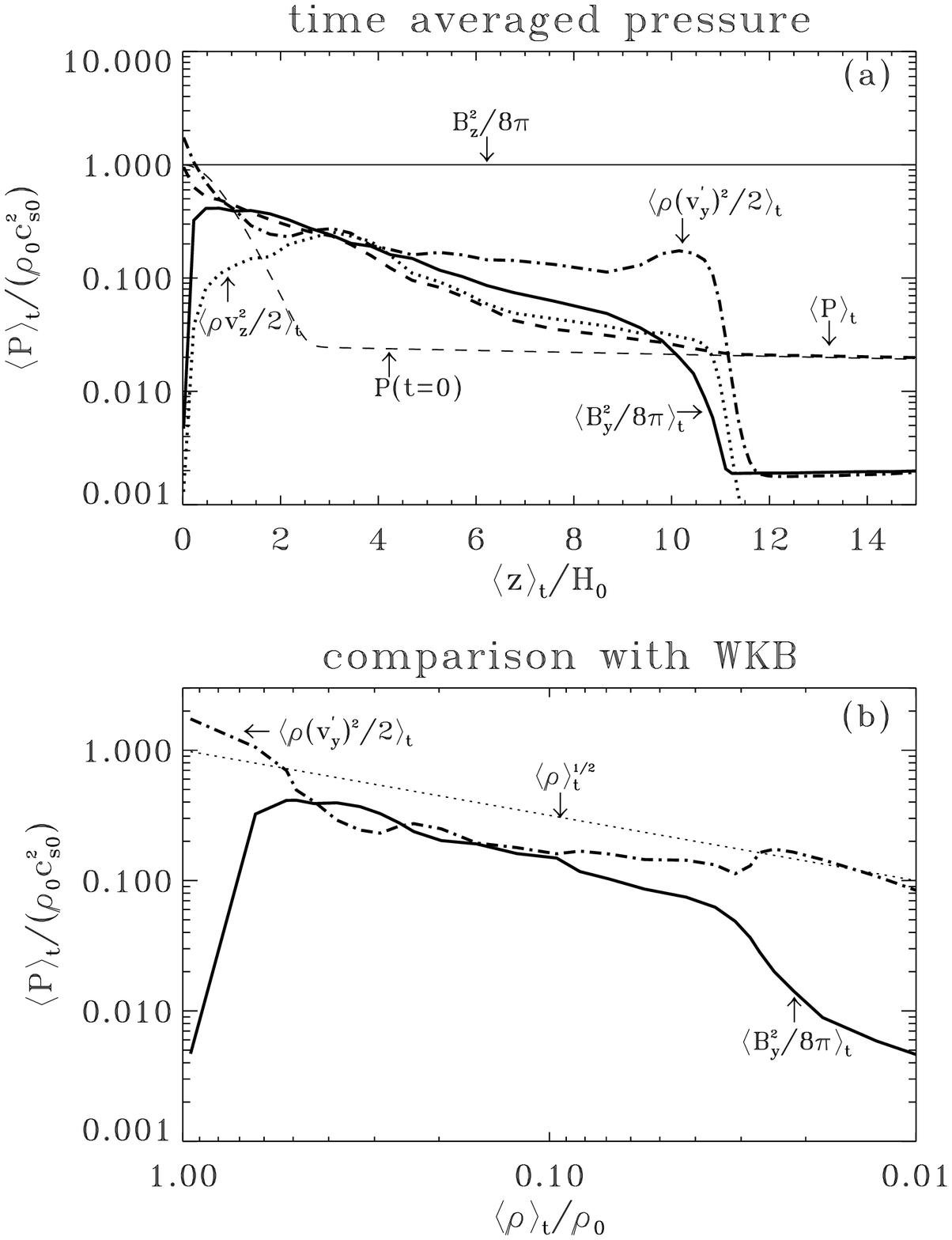,width=8.5cm}
\small
\smallskip\noindent {\sc Fig. ~7.~}\
(a) Time averages of thermal pressure, magnetic pressure,
and dynamic pressures for each Lagrangian fluid element as
a function of the time averaged position of each element.
The thick dashed line shows the thermal pressure, the thick solid
line shows the magnetic pressure of the $y$-component
of the magnetic field, the dash-dotted line shows the dynamic
pressure of the $y$-component of velocity, and the dotted line
shows the dynamic pressure of the $z$-component of velocity.
The thin solid line shows the magnetic pressure of background
magnetic field, and the thin dashed line shows the
initial thermal pressure.
(b) The dynamic pressure for the $y$-component of velocity (dash-dotted line)
and the magnetic pressure of the $y$-component of the magnetic field
(solid line) as a function of density.
The dotted line shows the scaling expected from the WKB theory.
\normalsize
\bigskip

In contrast to the WKB theory, the dynamic pressure of the
$z$-component of the velocity is nonzero in our 
results, although it is smaller than the dynamic pressure of
the $y$-component of the velocity.
According to the small-amplitude theory, the dynamic pressure of the 
$y$-component of the velocity and the magnetic pressure obey 
equipartition, i.e., 
\begin{equation}
\frac{1}{2}\rho (v^\prime_y)^2 = \frac{B_y^2}{8\pi} .
\end{equation}
Figure 7a shows that such an equipartition is almost satisfied between
$z=1H_0$ and $z=5H_0$, although there are deviations
near the midplane. This deviation comes from the effect 
of a nonzero driving force at $z=0$ and the symmetrical boundary condition 
$B_y(z=0)=0$. For $z > 5H_0$, a region which contains a small fraction of
the total cloud mass (see Figure 4), the two energies have distinct
spatial profiles. Near the interface between the cold and hot material 
(the cloud boundary), the magnetic pressure decreases more rapidly 
than expected from the WKB theory. We believe this is due primarily 
to a standing wave that is set up in the outer cloud, with the cloud boundary 
acting as a node for $B_y$ and an antinode for $v^\prime_y$. 
Although many different wave modes are generated by the turbulence
in the cloud, only those which satisfy the boundary condition for
a standing wave will interfere constructively upon reflection. 
A transverse standing wave can be set up even though the boundary itself is
moving, since the \Alf speed is much greater in the outer cloud than
the vertical speed of the boundary.
The influence of the boundary reaches well inside the cloud, 
since the effective wavelength of \Alf waves (scaling as $\rho^{-1/2}$
for waves of fixed frequency) is also quite large in the
low density region. For example, at $z= 5H_0$, where $\rho \simeq
0.1 \rho_0$, \Alf waves of the input frequency $\nu_0 = c_{s0}/H_0$ have 
wavelength $\lambda \simeq 4.5 H_0$; at $z= 10H_0$, it increases to 
$\lambda \simeq 8 H_0$.
An important result is that even though the waves are quasilinear in the
outer cloud ($B_y \ll B_0$), the transverse velocity amplitude
is much larger than expected from equipartition arguments.

In Figure 7b, we show the dynamic pressure of 
the $y$-component of velocity and the magnetic pressure of 
the $y$-component of the magnetic field
as a function of density in order to compare with 
the WKB model predictions. 
We find that the energy in transverse 
velocities, $\onehalf \rho (v^\prime_y)^2$, scales approximately
as $\rho^{1/2}$ in the outer cloud, the same as the WKB theory, 
although there is a noticeable upward turn at $z \simeq 9 H_0$ due to
the standing wave effect. However, the wave magnetic energy 
$B_y^2/(8 \pi)$, which is directly responsible for vertical support, 
decreases more rapidly than in the WKB model in the 
outer cloud, scaling approximately as $\rho$.
We note that strict adherence to the WKB predictions is not
expected in our model due to nonlinearity,
wave dissipation, and the low frequency $\nu_0=c_{s0}/H$ 
of the input turbulence.

\subsubsection{Energies in the Cloud}

Figure 8 shows the time evolution of various energies in the cloud.
These are calculated as follows:
kinetic energy of the $z$-component of velocity,
\begin{equation}
E_{kz} (t) = \int_0^{z_f(t)} \frac{1}{2} \rho v_z^2 dz;
\label{eq:ekz}
\end{equation}
kinetic energy of the $y$-component of velocity,
\begin{equation}
E_{ky} (t) = \int_0^{z_f(t)} \frac{1}{2} \rho v_y^2 dz 
-E_{km},
\label{eq:eky}
\end{equation}
where $E_{km}$ is the kinetic energy of the mean motion of the cloud, i.e.,
\begin{equation}
E_{km}(t)= \frac{1}{2} \langle v_m\rangle^2 \int_0^{z_f(t)} \rho dz;
\end{equation}
magnetic energy of the $y$-component of magnetic field,
\begin{equation}
E_{m} (t) = \int_0^{z_f(t)} \frac{B_y^2}{8\pi}  dz;
\label{eq:em}
\end{equation}
sum of the above terms,
\begin{equation}
E_{T} (t) = E_{kz} (t) + E_{ky} (t) + E_{m} (t).
\label{eq:et}
\end{equation}
In the above equations, the integration was done from 0 to $z_f(t)$ 
because we are interested in energies of the cold material.
Strictly speaking, the total energy in each case is twice the value
we calculate due to the mirror symmetric boundary condition at $z=0$.
Equation (\ref{eq:eky}) shows that the mean kinetic energy of the 
cloud is subtracted from the kinetic energy of $y$-component of velocity,
so that $E_{ky}$ means the kinetic energy of the oscillating velocity.

\bigskip
\psfig{file=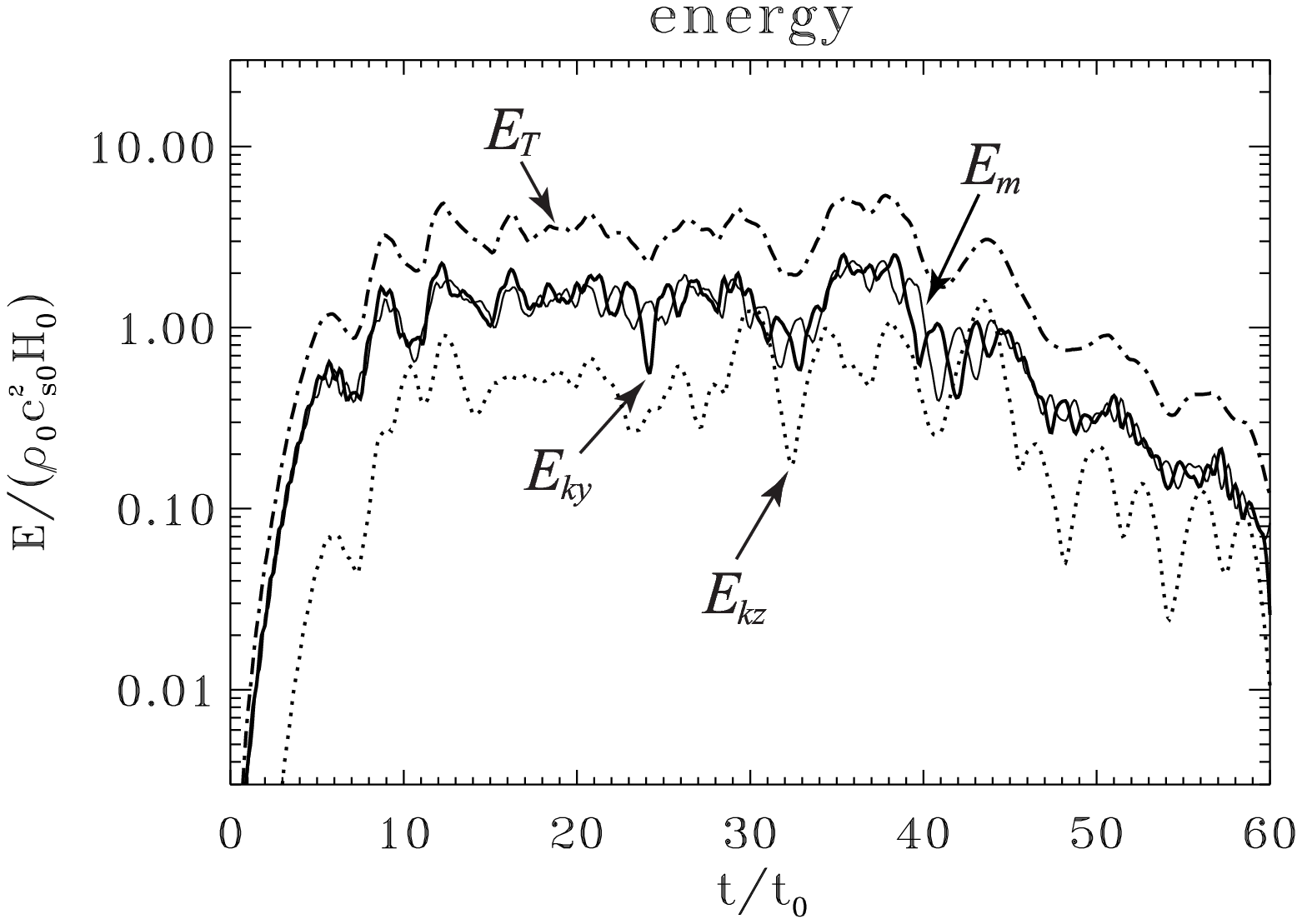,width=8.5cm}
\small
\smallskip\noindent {\sc Fig. ~8.~}\
Time evolution of various energies in the cloud.
The thick solid line shows $E_{ky}$, the dotted line shows $E_{kz}$,
the thin solid line shows $E_{m}$, and dash-dotted line shows $E_{T}$.
The values of each energy are smoothed out over every one cycle of
the driving force in order to remove periodic oscillations originating from
the driving force.
\normalsize
\bigskip

In Figure 8, the thick solid line shows $E_{ky}$, the dotted line 
shows $E_{kz}$, the thin solid line shows $E_{m}$, and dashed-dotted 
line shows $E_{T}$. 
The values of each energy are smoothed out over every one cycle of 
the driving force in order to remove periodic oscillations originating from
the driving force. Among the energies, $E_{ky}$ and $E_{m}$ are 
comparable to each other, but $E_{kz}$ is significantly smaller 
than the others. The sum of energies
$E_{T}$ is almost constant in a logarithmic scale 
while the input energy is constant ($t=10t_0-40t_0$), 
though it has some fluctuations. After the driving force is terminated
at $t=40t_0$, the energies decrease almost exponentially.
The decreasing time is $\simeq 8.5t_0$. 

\subsection{Parameter Dependence on the Strength of the Driving Force}

\subsubsection{Density, Velocity and Pressure}

In this paper, we study the effect of changing the strength 
of the driving force, by changing $a_d$ in equation (\ref{eq:driving2}).

Figure 9 shows the time evolution of densities for $\tilde{a}_d=20$ 
and $\tilde{a}_d=40$. The $y$-velocities at $z=0$, $v^\prime_y$,
are also shown in the figure. This figure shows that a stronger driving 
force causes a larger turbulent velocity, which results 
in a more dynamic evolution of the molecular cloud, 
including stronger shock waves, and larger excursions of the cloud
boundary. However, after terminating the driving force at $t=40t_0$,
the shock waves dissipate and the clouds shrink in both cases.

\bigskip
\psfig{file=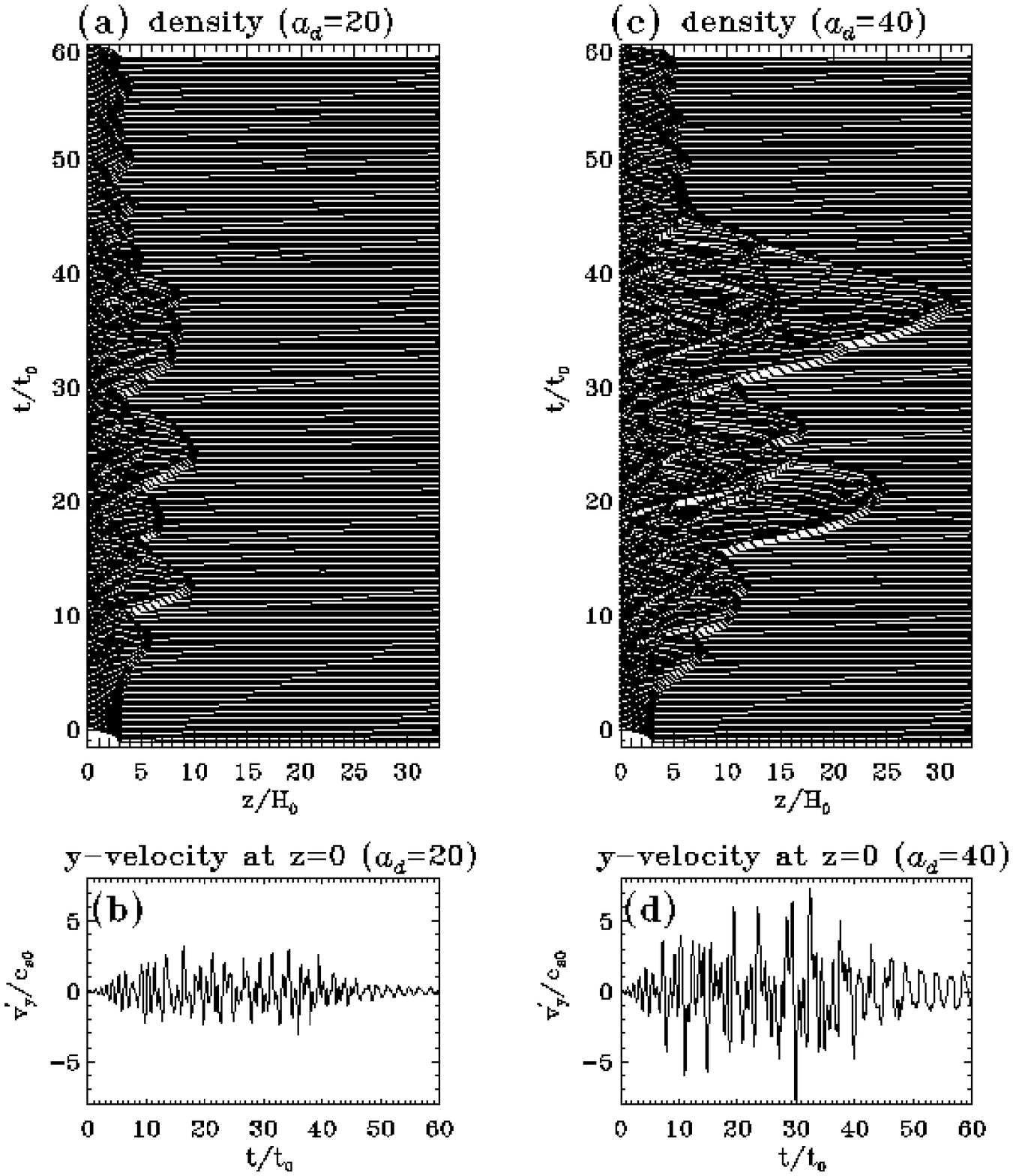,width=8.5cm}
\small
\smallskip\noindent {\sc Fig. ~9.~}\
Evolution of clouds with $\tilde{a}_d=20$ and $\tilde{a}_d=40$.
(a) Time evolution of densities for $\tilde{a}_d=20$.
(b) The oscillating part of $y$-velocity at $z=0$, $v^\prime_y$,
as a function of time for $\tilde{a}_d=20$.
(c) Time evolution of densities for $\tilde{a}_d=40$.
(d) The oscillating part of $y$-velocity at $z=0$, $v^\prime_y$,
as a function of time for $\tilde{a}_d=40$.
The density plots at various times are stacked with time
increasing upward in uniform increments of $0.2t_0$.
\normalsize
\bigskip

Figure 10 shows the time average of the density
and velocities for each Lagrangian fluid element 
for $\tilde{a}_d=20$ and $\tilde{a}_d=40$.
The time average is taken between $t=10t_0$ and $t=40t_0$.
This figure also shows that a stronger driving force causes 
the cloud to move further outward and a 
larger velocity dispersion within the cloud.

\bigskip
\psfig{file=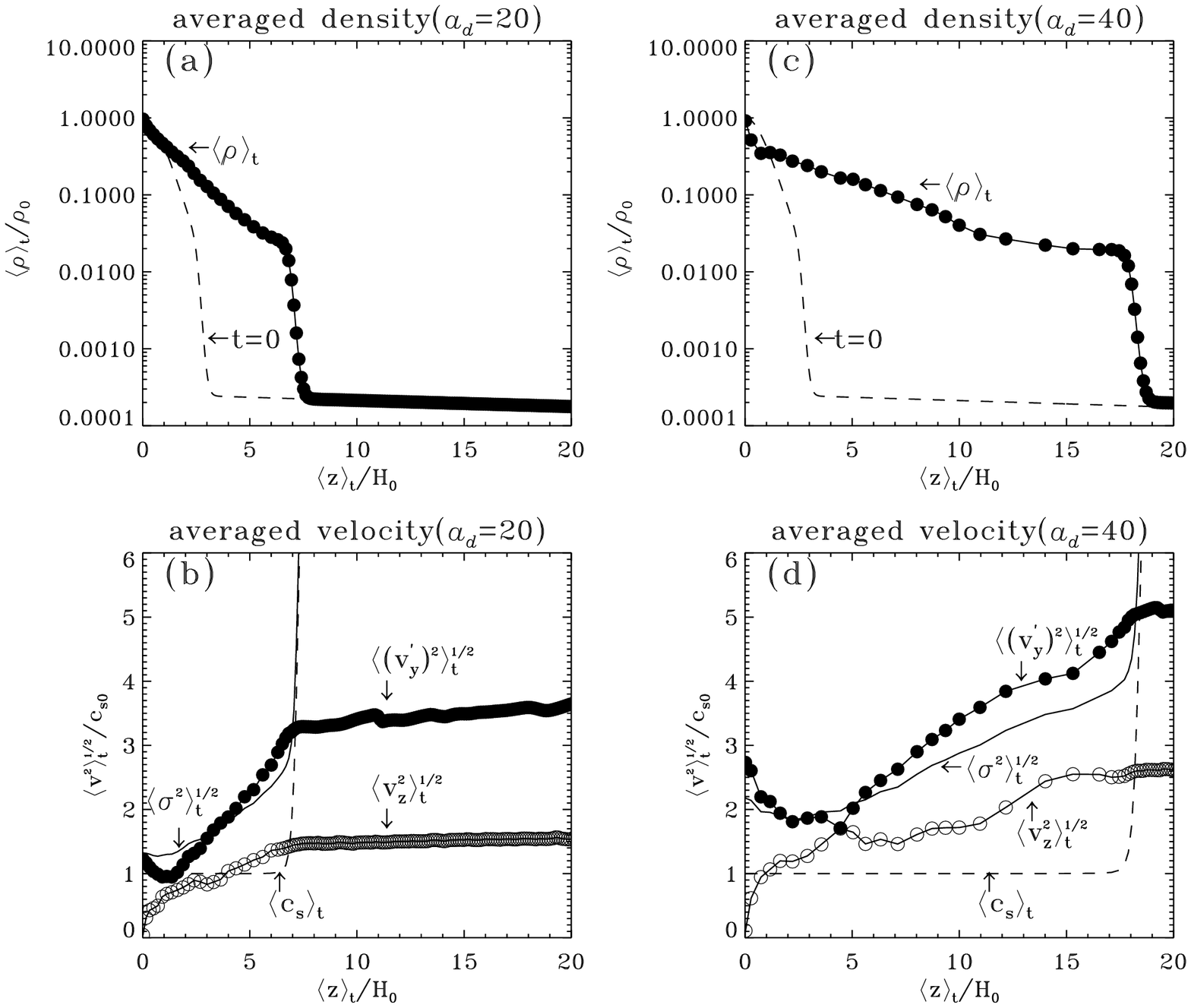,width=8.5cm}
\small
\smallskip\noindent {\sc Fig. ~10.~}\
Density and velocity dispersions for the cases $\tilde{a}_d=20$ and
$\tilde{a}_d=40$.
(a) Time average of the density for each element
as a function of the time averaged position of each element
for $\tilde{a}_d=20$.
The time average is calculated between $t=10t_0$ and $t=40t_0$.
Each Lagrangian fluid element is equally spaced
at time $t=0$ with spacing $\Delta z =0.1H_0$ starting
at $z=0.01 H_0$.
(b) The time averaged root mean square of various velocities for
each Lagrangian fluid element plotted versus
the time average of the position of each element
for $\tilde{a}_d=20$.
Open circles show the time averaged root mean square of $v_z$,
filled circles show that of $v^\prime_y$,
and the dashed line shows the time average of the sound speed $c_s$.
The solid line shows the root mean square of the time average
of the effective one-dimensional velocity dispersion.
(c)  The time average of the density for each element
as a function of the time averaged position of each element
for $\tilde{a}_d=40$.
(d) The time averaged root mean square of various velocities for
each Lagrangian fluid element plotted versus the time average of
the position of each element for $\tilde{a}_d=40$.
\normalsize
\bigskip

Figure 11 shows the time average of pressures
for $\tilde{a}_d=20$ and $\tilde{a}_d=40$.
When the driving force is strong, the magnetic pressure
and dynamic pressure of the $z$-component of velocity
become significantly larger than the thermal pressure.

\bigskip
\psfig{file=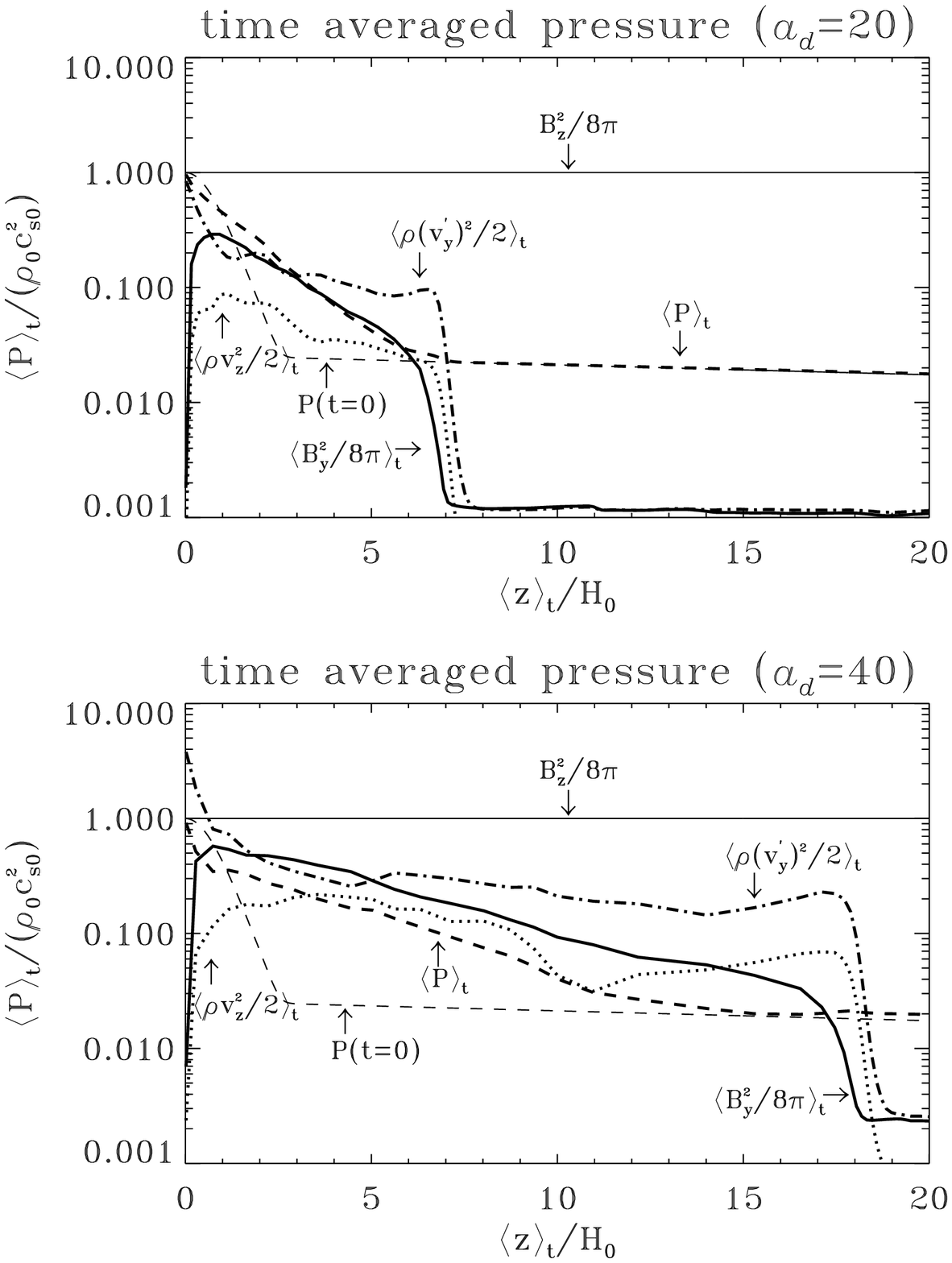,width=8.5cm}
\small
\smallskip\noindent {\sc Fig. ~11.~}\
Time averages of the thermal pressure, magnetic pressure,
and dynamic pressures for each Lagrangian fluid element as
a function of the time averaged position of each element.
The upper panel is for $\tilde{a}_d=20$ and the lower panel is
for $\tilde{a}_d=40$.
The thick dashed line shows the thermal pressure,
the thick solid line shows the magnetic pressure of the $y$-component
of the magnetic field, the dash-dotted line shows the dynamic pressure
of the $y$-component of velocity, and the dotted line shows the dynamic
pressure of the $z$-component of velocity.
In both panels, the thin solid line shows the magnetic pressure
of background magnetic field, and the thin dashed line shows the
initial thermal pressure.
\normalsize
\bigskip

\subsubsection{Energy}

Figure 12 shows the time evolution of $E_{T}$ for both 
$\tilde{a}_d=20$ and $\tilde{a}_d=40$.
These values are also smoothed out over every one cycle of 
the driving force to remove periodic oscillations. 
In the case of $\tilde{a}_d=20$, the energy is almost constant 
in a logarithmic scale while the driving amplitude is constant 
($t=10t_0-40t_0$).
However, in the case of $\tilde{a}_d=40$, the energy is still gradually 
increasing 
until $t=30t_0$, but afterwards becomes almost constant until $t=40t_0$.
After terminating the driving force at $t=40t_0$,
both energies decrease almost exponentially.
The energy decreasing times, $t_d$, for each parameter, 
which are estimated by fitting an exponential function, 
are listed in Table 1. 
In this study, we terminate the driving force at $t=40t_0$ in every
case for simplicity. However, we found that the energy decreasing
time can vary somewhat depending on when we terminate the driving force. 
Accounting for this as well as the fitting error of the exponential function,
we conclude that the energy decreasing times have a range of variation
about $\pm 2t_0$ of the values listed in Table 1. 

\bigskip
\psfig{file=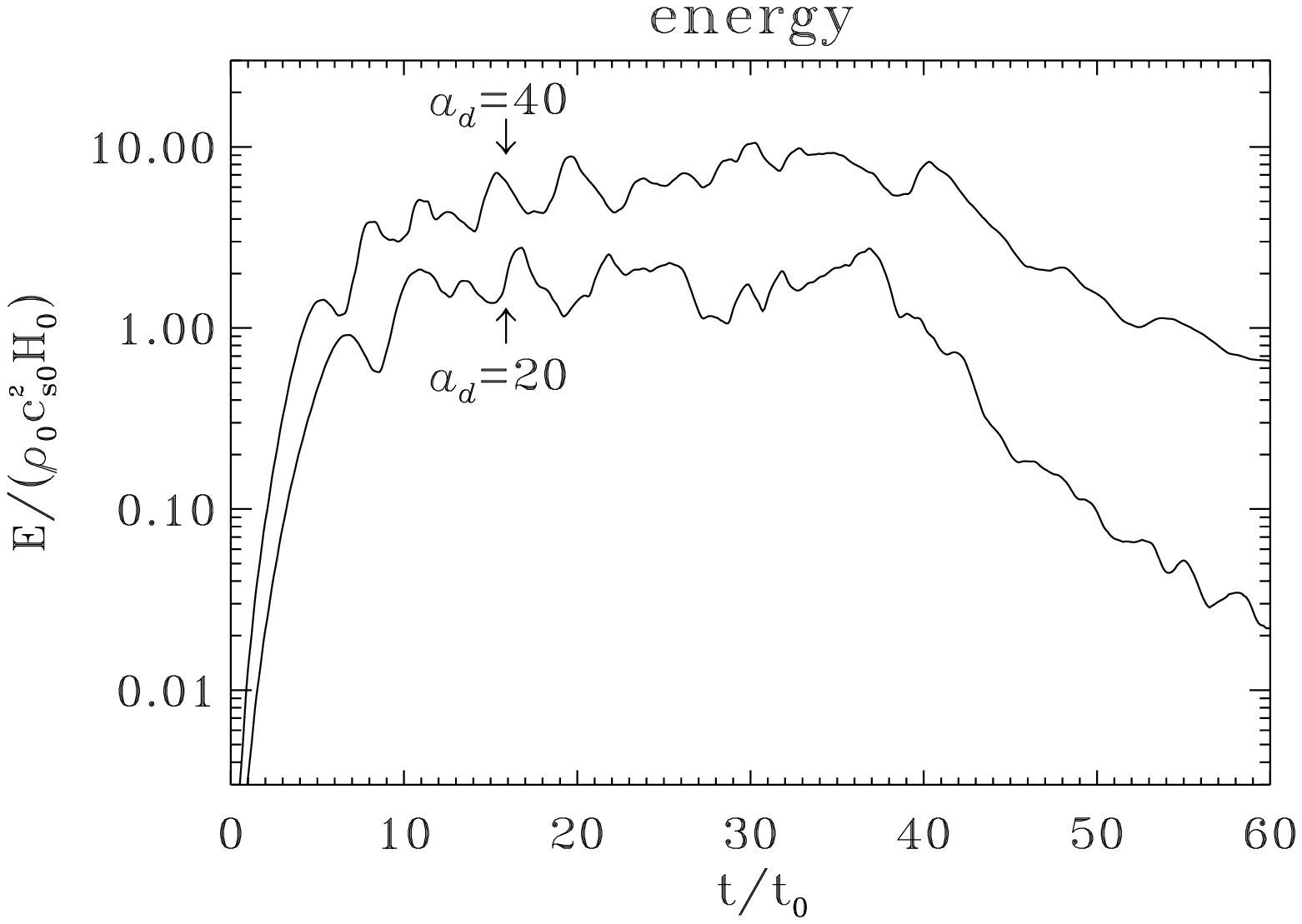,width=8.5cm}
\small
\smallskip\noindent {\sc Fig. ~12.~}\
Time evolution of $E_{T}$ for both
$\tilde{a}_d=20$ and $\tilde{a}_d=40$.
The values of each energy are smoothed out over every one cycle of
the driving force in order to remove periodic oscillations originating from
the driving force.
\normalsize

\subsubsection{Correlations between Velocities and Sizes}

Here we investigate the correlation between 
the velocity dispersion and the height of the cloud.
Figure 13a shows the time averaged velocity dispersions 
$\langle \sigma^2 \rangle_t^{1/2}$ of 
different Lagrangian fluid elements for different $\tilde{a}_d$, 
as a function of $\langle z \rangle_t$.
The open circles correspond to Lagrangian fluid elements
whose initial positions are located at $z=2.51$,
which is close to the edge of the cold cloud.
The filled circles corresponds to Lagrangian fluid elements
whose initial positions are located at $z=0.61$,
which is approximately the half-mass position of the cold cloud.
Each circle corresponds to a different value of $\tilde{a}_d$.
These values are summarized in Table 1.
The dotted line shows  
\begin{equation}
\langle \sigma^2 \rangle_t^{1/2} \propto \langle z \rangle_t^{0.5}.
\end{equation}
This figure shows that the velocity dispersions have 
a good correlation with the heights of the molecular clouds.

\bigskip
\psfig{file=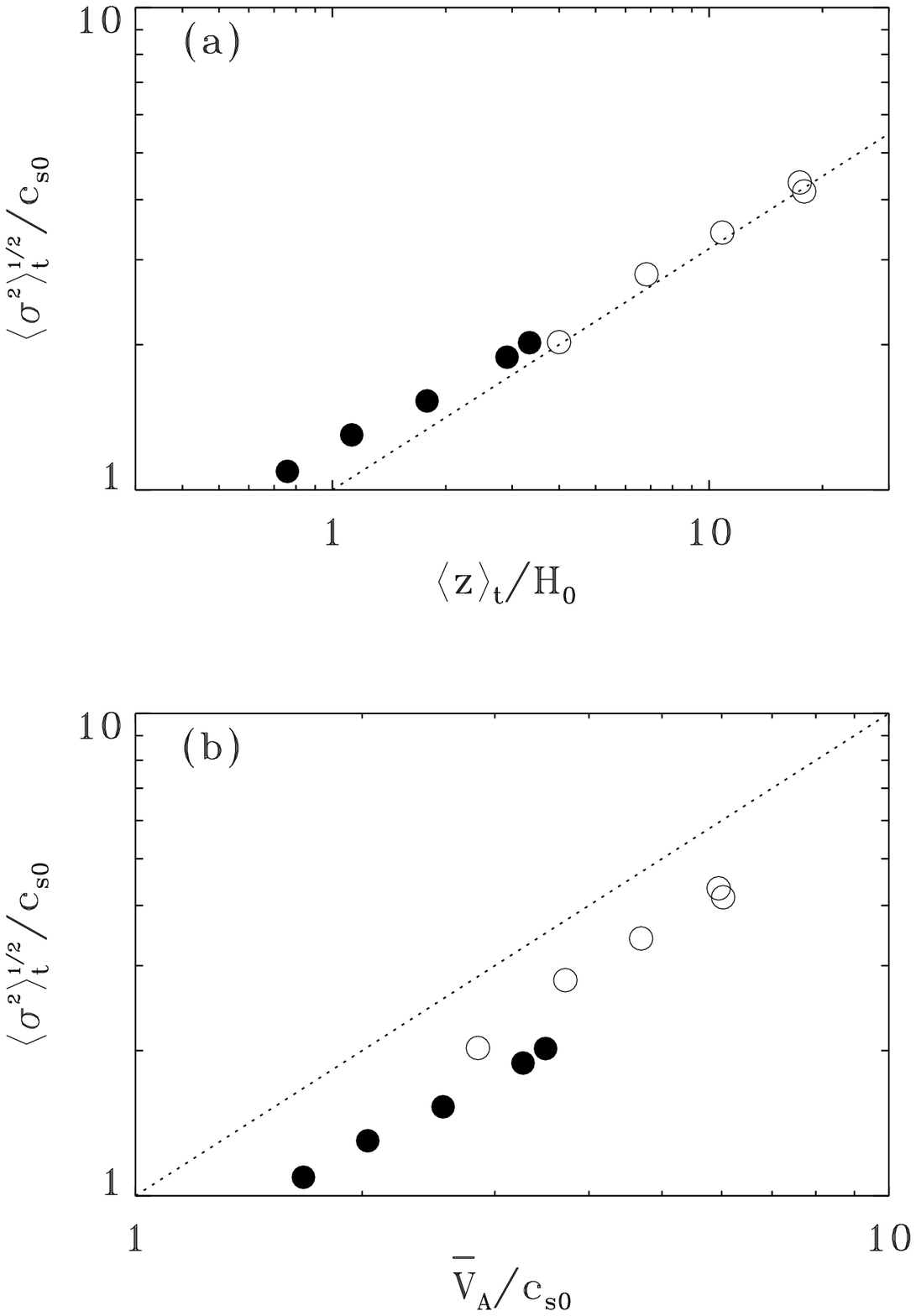,width=8.5cm}
\small
\smallskip\noindent {\sc Fig. ~13.~}\
Global properties of an ensemble of clouds with different driving
strengths $\tilde{a}_d$.
(a) Time averaged velocity dispersions of
different Lagrangian fluid elements for different $\tilde{a}_d$,
as a function of time averaged positions.
The open circles correspond to Lagrangian fluid elements
whose initial positions are located at $z=2.51$,
which is close to the edge of the cold cloud.
The filled circles correspond to Lagrangian fluid elements
whose initial positions are located at $z=0.61$,
which is approximately the half-mass position of the cold cloud.
The dotted line shows
$\langle \sigma^2 \rangle_t^{1/2} \propto \langle z \rangle_t^{0.5}$.
Each circle can be associated with a particular model in our study by
comparison with the numbers in Table 1.
(b) Time averaged velocity dispersions as a function of
the mean Alfv\'en velocity of the cloud.
The dotted line shows
$\langle \sigma^2 \rangle_t^{1/2} \propto \bar{V}_A$.
\normalsize
\bigskip

Figure 13b shows the correlation between the velocity dispersion
and mean Alfv\'en velocity of the cloud $\bar{V}_A$ at the mean position
$\langle z \rangle_t$ of the Lagrangian elements. 
The dotted line shows  
\begin{equation}
\langle \sigma^2 \rangle_t^{1/2} \propto \bar{V}_A.
\end{equation}
This figure shows that the velocity dispersions have 
a good correlation with the mean Alfv\'en velocity, defined by
\begin{equation}
\bar{V}_A \equiv \frac{B_0}{\sqrt{4\pi\bar{\rho}}},
\label{eq:mva}
\end{equation}
where 
\begin{equation}
\bar{\rho}=\frac{\Sigma}{2 \langle z \rangle_t}
\label{eq:mde}
\end{equation}
is the mean density and $\Sigma$ is the column density for each 
Lagrangian element.
We discuss the meaning of these correlations in the next section.

\section{Discussion}

\subsection{Velocity Dispersion and Equilibrium}

In Figure 13a, we found that the velocity dispersion obeys the linewidth-size 
relation 
\begin{equation}
\langle \sigma^2 \rangle_t^{1/2} \propto \langle z \rangle_t^{0.5}.
\label{eq:lwsizesim}
\end{equation} 
This relation is satisfied (at both the full-mass and half-mass Lagrangian 
positions) for an ensemble of the clouds with different strengths of 
the driving force. It is consistent with observational results of molecular 
clouds (Larson 1981; Myers 1983; Solomon et al. 1987).
Within any individual cloud, $\langle \sigma^2 \rangle_t^{1/2}$ also
increases toward the cloud boundary.

A similar but not identical relation, relating the velocity dispersion 
at $z=0$ to the 
size of the cloud, is expected in virtually any one-dimensional model,
regardless of the form of spatial variation of the pressure within the
cloud. This is because an integral over the vertical force-balance equation 
reveals that the total pressure at $z=0$, $P_{\rm tot,0}$ must equal the weight
of the accumulated gas above, $\pi G \Sigma^2/2$, assuming that the
surface pressure is negligible. If $P_{\rm tot,0} = \rho_0 
\sigma_{\rm eff,0}^2$,
where $\sigma_{\rm eff,0}$ is the effective velocity dispersion at $z=0$, and
$\Sigma = 2 \rho_0 Z$, in which $Z$ is the typical size scale of the cloud,
then one naturally obtains 
\begin{equation}
\sigma_{\rm eff,0} \propto Z^{1/2}
\label{eq:lwsizegen}
\end{equation}
for an ensemble of clouds of fixed total column density $\Sigma$
but varying $\sigma_{\rm eff,0}$.
This relation applies to the Spitzer equilibrium state, in which 
$\sigma_{\rm eff,0} = c_s$ and $Z=H_0$. It also applies to the 
equilibrium state calculated by Martin et al. (1997), in which 
$\sigma_{\rm eff,0}$ is the effective velocity dispersion associated with
\Alf waves at $z=0$.  While a similar relation will inevitably apply to
our nonlinear model as well, we note that the relation (\ref{eq:lwsizesim})
is more general, in that it relates the velocity dispersion {\it at}
the position (not $z=0$) of a Lagrangian mass element to the position of 
that element. 
This more general correlation is a less predictable property of our 
time-averaged equilibrium state. It is also closer to what is 
often measured, since optical depth effects may mean that 
a measured velocity dispersion samples the largest scales of an observed 
cloud rather than the center.

Equation (\ref{eq:lwsizesim}) is strongly related to 
another relation 
\begin{equation}
\langle \sigma^2 \rangle_t^{1/2} \propto \bar{V}_A
\label{eq:lwva}
\end{equation}
which is shown in Figure 13b, once more for both the half-mass and
full-mass positions for our ensemble of clouds.
This relation is also consistent with 
observational results of molecular clouds (Crutcher 1999; Basu 2000).
For example, Figure 1b of Basu (2000) shows an excellent correlation
between the line-of-sight component of the large-scale magnetic field 
$B_{\rm los}$ and $\sigma \rho^{1/2}$ for observed clouds of widely varying
values of $\rho$, $B_{\rm los}$, and $\sigma$, essentially the same
correlation as equation (\ref{eq:lwva}) if $B_{\rm los} \propto B$ on average.

Here we would like to point out that the relation (\ref{eq:lwva})
does not necessarily imply that turbulent motions are due to 
\Alf waves, although that is primarily the case in our simulation.
The relation is far more general in that it is obtained 
when the velocity dispersion is caused by any
mechanism which results in the clouds being in virial balance
between gravitational and turbulent energies, as well as having
a mass-to-flux ratio that is approximately uniform from cloud to cloud
(Mouschovias 1987; Shu et al. 1987). 
Rather than imagining a time-independent mean \Alf speed in a cloud
(strictly true in an incompressible medium or a periodic box
simulation with a fixed mean density), and turbulent motions 
becoming comparable to that speed, our simulation reveals that it is the
mean \Alf speed of any cloud which readjusts to a new value as $\bar{\rho}$
drops to accommodate a particular level of turbulent driving. 
In this view, the turbulent dispersion $\sigma$ is the more 
fundamental quantity, dependent on the particular source of
driving, and $\bar{V}_A$ is a quantity which readjusts
to become comparable to $\sigma$.

Finally, we note that 
the relation (\ref{eq:lwva}) depends on the definition 
of the mean Alfv\'en velocity.
If we use the local Alfv\'en velocity of each particle,
we could not get a clear correlation like equation (\ref{eq:lwva}).
The mean Alfv\'en velocity is often calculated observationally using
the mean density, as in equations (\ref{eq:mva}) and (\ref{eq:mde}),
because it is difficult to directly measure a local Alfv\'en velocity.
The local Alfv\'en velocity in a gravitationally stratified cloud
often takes on values much different than the overall mean quantity
$\bar{V}_A$ used in equation (\ref{eq:lwva}). 
For example, the time averaged local Alfv\'en velocity near the
edge of the cloud is about 2.6 times greater than $\bar{V}_A$
for the case of $\tilde{a}_d=30$.

\subsection{Relation to Linear Model and Chandrasekhar-Fermi Formula}

In our simulation, the velocity component parallel to the 
background magnetic field ($v_z$) is generated by the nonlinear 
effect of the waves. However, the time-averaged magnitude of $v_z$ 
is significantly less than the time-averaged magnitude of $v_y^\prime$
in all our models (see Figures 6, 7, 10, and 11).
Although the waves are nonlinear, the coupling between the
\Alf and slow mode MHD waves is not so strong as to destroy an
approximate equipartition between
$B_y^2/(8\pi)$ and $\onehalf \rho (v_y^\prime)^2$ throughout 
much of the cloud, which is the expected result for purely transverse
linear \Alf waves.  However, this equipartition does break down in
the outer part of the
cloud, where standing wave motions are established in which
$B_y$ has a node at the cloud boundary, and $v_y^\prime$ has an antinode.

The breakdown of the $v_y^\prime$ v.s. $B_y$ relation in outer cloud
means that the original Chandrasekhar-Fermi formula breaks down there 
as well.
If we assume that the dispersion of polarization angle of the magnetic field 
($\delta \theta$) is related to the velocity dispersion as
\begin{equation}
\delta \theta = \frac{|B_y|}{B_0} = \alpha \frac{|v_y^\prime|}{V_A},
\end{equation}
we can estimate the strength of the magnetic field of the cloud 
by using the observational values of polarization angle,
density, and velocity dispersion. This yields
\begin{equation}
B_0= \alpha (4\pi\rho)^{1/2} |v^\prime_y| (\delta \theta)^{-1},
\label{eq:chfe}
\end{equation}
where $\alpha$ is a nondimensional factor and 
equals 1 for linear Alfv\'en waves.
The relation (\ref{eq:chfe}) with $\alpha = 1$ was proposed by 
Chandrasekhar \& Fermi (1953) in order to estimate the strength
of the background magnetic field in the interstellar medium.
The generalized form with $\alpha \neq 1$
can be fit to our simulation if we define 
$\alpha$ in our simulation as the square root of the ratio of 
time averaged magnetic pressure to dynamic pressure
for each Lagrangian fluid element, i.e., 
\begin{equation}
\alpha=
\left( 
\frac{\langle B_y^2/(8\pi) \rangle_t}
{\langle \rho (v^\prime_y)^2/2 \rangle_t}
\right)^{1/2}.
\end{equation}
The resulting distribution of $\alpha$ as a function
of $\langle z \rangle_t$ for the case of $\tilde{a}_d=30$ is shown in
Figure 14. The time averaged density, which was shown in Figure 4,
is also shown as a dashed line to clarify the edge of
the cloud. 
It shows that $\alpha$ is close to unity inside the cloud in the region
where most of the mass is enclosed (see Figure 4), 
but it decreases near the surface of the cold cloud, where the
standing wave effect becomes important;
the minimum value is about $\alpha=0.23$.
Outside of the cloud ($\langle z \rangle_t > 12 H_0$ in Figure 14),
$\alpha=1$ since the waves are linear there due to the low density and
high ambient \Alf speed. 
Therefore, while our nonlinear model supports the use of the 
Chandrasekhar-Fermi formula throughout most of a stratified cloud, 
we caution against its use with $\alpha = 1$ near the surface of a cloud.

\bigskip
\psfig{file=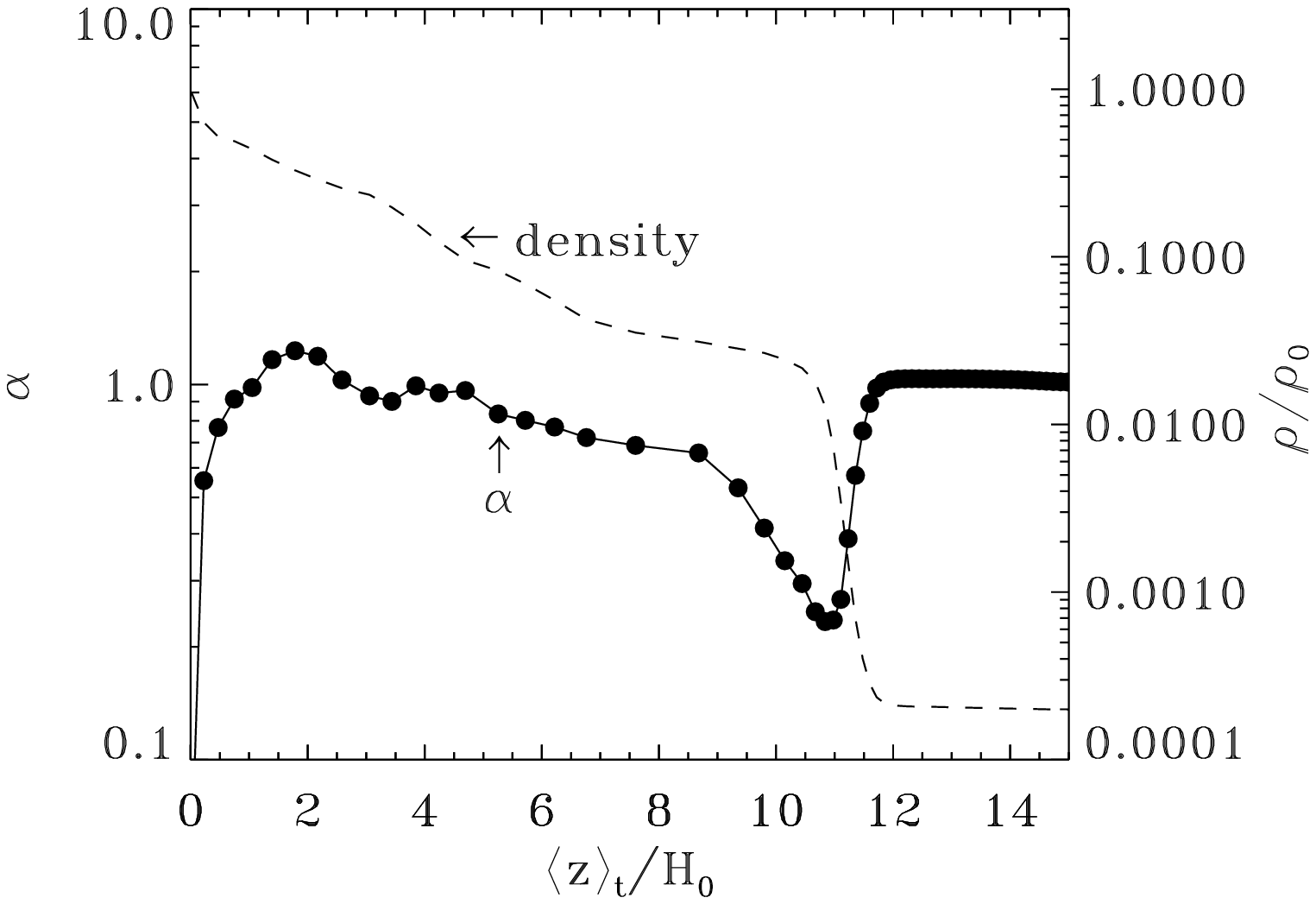,width=8.5cm}
\small
\smallskip\noindent {\sc Fig. ~14.~}\
Parameter $\alpha$ which appears in the Chandrasekhar-Fermi formula
(eq. [\ref{eq:chfe}]),
as a function of the time averaged position of fluid elements within the cloud
for the case of $\tilde{a}_d=30$.
The time averaged density, which was shown in Figure 4,
is also shown as a dashed line, mainly to clarify the position
of the edge of the cloud.
\normalsize

\subsection{Global Oscillations}

Figure 15 shows the time evolution of the position of a Lagrangian 
fluid element whose initial position is $z=2.51$, for the standard 
model with driving strength $\tilde{a}_d=30$.
This corresponds to the motion of the outer edge of the cloud.
The motions resemble longitudinal normal mode oscillations, 
with the excursions of the outer cloud very similar to free-fall. 
The dotted line in the figure shows the trajectory for
free-fall motion of the fluid element for several different time intervals.
The peak of the trajectories are chosen to coincide with a peak of the 
oscillation in the computed model. The paths are
parabolic due to the constant gravity acting on a comoving mass shell in 
the one-dimensional approximation. 
Similar to the transverse standing wave that is set up in the outer
cloud, the longitudinal motions in this region also resemble a standing wave
pattern with an antinode of $v_z$ at the cloud boundary. The outermost
part of the cloud suffers the greatest displacements, as in the case
of a pulsating star.
Once the internal driving is discontinued ($t > 40 \, t_0$), 
the outer surface also moves inward in nearly a free-fall manner.

\bigskip
\psfig{file=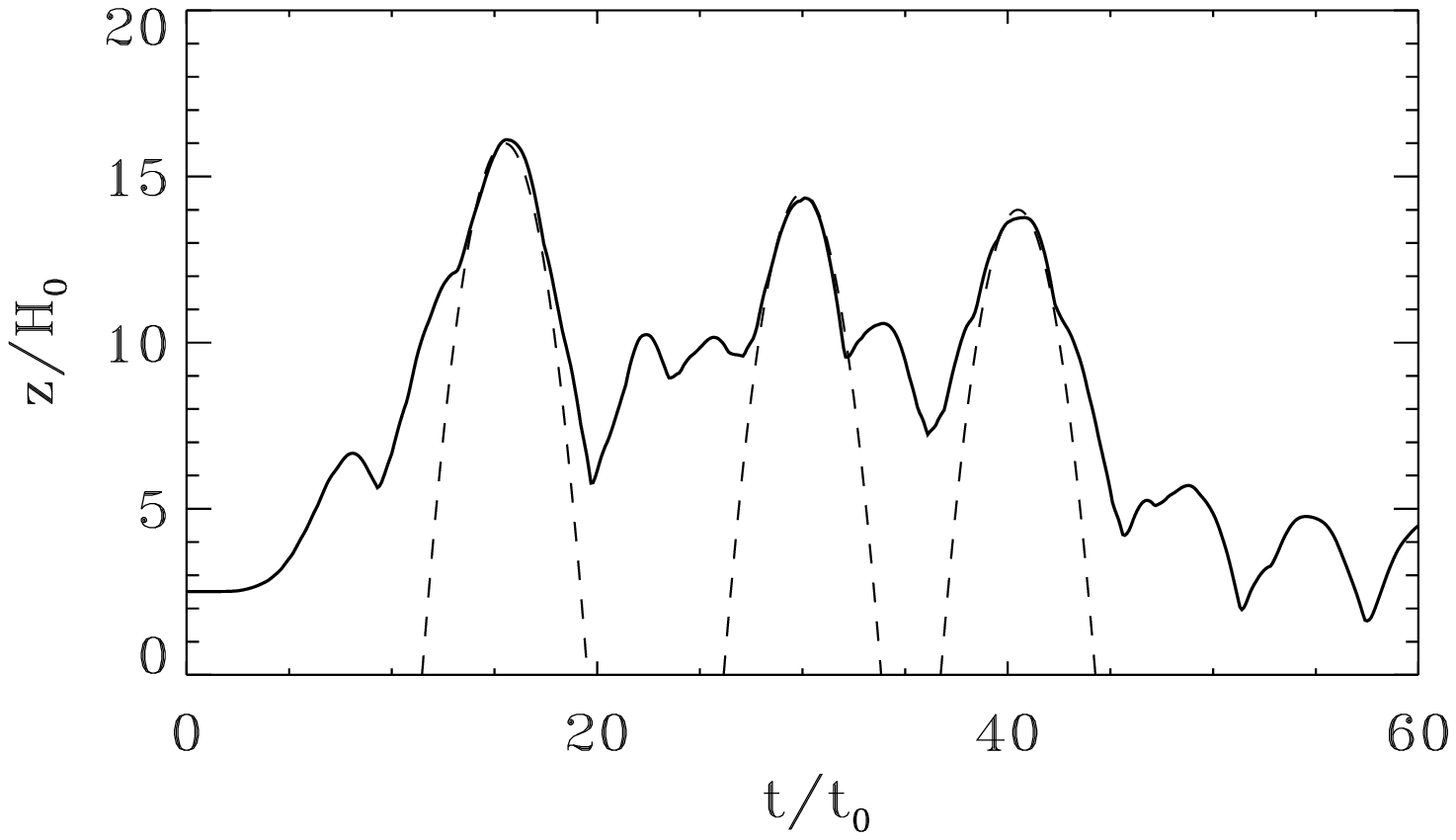,width=8.5cm}
\small
\smallskip\noindent {\sc Fig. ~15.~}\
Time evolution of the position of a Lagrangian
fluid element whose initial position is located at $z=2.51$
for the case of $\tilde{a}_d=30$.
The dotted line shows the trajectories of free-fall
motion for the element.
\normalsize
\bigskip

Figure 3 and Figure 15 also show that some residual oscillations remain 
on the largest scale even after the damping of most of the internal
turbulent energy. This effect is reminiscent of the recent
observation of Lada et al. (2003) that implied a global oscillation
of Barnard 68, a cloud with only subsonic internal turbulence. 
We hypothesize that Barnard 68 may have dissipated most of the internal 
turbulence left over from its formation,
but that only a large-scale oscillation remains, as in the late stage
of our simulation after turbulent driving is discontinued.
However, we caution that ours is a one-dimensional model, and 
globally coherent standing-wave motions need to be demonstrated in 
a multi-dimensional model.

\subsection{Dissipation of Energy}

The dissipation of energy that we input into nonlinear transverse \Alf modes
is caused though either shock waves or grid-scale 
dissipation, which is similar to the results of Stone et al. (1998). 
In addition to the dissipation, a part of the energy escapes from 
the cloud in our simulation. 
We measured the Poynting flux, which is the dominant energy
flux, at the full-mass position and the 20-percent mass position,
which is just outside the region we input the driving force.
The time integral of the Poynting flux at the full-mass position
is about 30 percent of that at the 20-percent mass position.

In our model, the majority of dissipation occurs due to the transfer
of energy to small scales via nonlinear steepening and/or a turbulent
cascade, followed by damping on
the grid scale though numerical resistivity. This is because
the kinetic energy of longitudinal motions,
which are the cause of shock waves, is smaller than that of transverse motions. 
The primary dissipation of Alfv\'enic motions then, is not due to 
coupling to slow modes, although this is present at a significant level.
This is similar to the finding of the one-dimensional numerical
model of Gammie \& Ostriker (1996), although the same authors
claim that in a multi-dimensional simulation, compressive effects
become a major source of dissipation (Stone et al. 1998); however, see
Cho \& Lazarian (2003), who find that the primary dissipation mechanism
in a three-dimensional simulation is not the coupling of \Alf modes 
to slow and fast modes. 

So if the primary source of dissipation in this simulation is grid-scale 
dissipation after nonlinear steepening of waves (and perhaps some cascade 
to smaller wavelengths), then is the dissipation rate a physical effect 
or an artifact of the simulations? To answer this, we have
increased the resolution of our numerical simulations
until the dissipation rate is nearly independent of grid size 
for the case of the typical parameter. 
Hence, the numerical diffusion 
is no longer resolution-determined at our current resolution of
50 points per initial scale height in the region where most of the 
mass is located.
If we use a smaller grid size, the final scale of the dissipation 
is smaller, but the rate of dissipation stays nearly the same.
In reality, there must be a physical effect such as ion-neutral
friction which would damp the waves at small scales.

The dissipation time of our results are a few crossing times of 
the time-averaged scales of the clouds (see Table 1).
They are a bit longer than estimated from periodic box simulations.
We think that this comes from the generation of longer-wavelength modes
as the waves travel to low density regions near the cloud's surface. 
However, it is already known that one-dimensional simulations have lower 
dissipation rates than two or three-dimensional ones (Ostriker et al. 2001). 
Therefore, higher dimensional global simulations will give the final 
answer in the future.

\subsection{Future Work}

This is the first in a series of papers.
In the next paper, we will conduct a complete survey of
the effect of important parameters such as $\beta_0$ and $\nu_0$.
Additionally, the study of random, rather than sinusoidal, disturbances
will be considered.
Moreover, we will study the case of circularly polarized Alfv\'en 
waves by including an $x$-component of motions in the simulation.
A circularly polarized wave is possibly less 
dissipative than a linearly polarized wave, since 
in the theoretical limit of an infinite wave train, a 
flux of circularly polarized waves has no associated magnetic pressure
gradient and thereby induces no compression of gas. Although conversion of
\Alf modes to slow modes is already not a dominant source of dissipation
in our model, it will be useful to see if the dissipation rate is even 
less than measured here when circularly polarized waves propagate in a 
finite-sized cloud.

We also expect to include ion-neutral friction in a future simulation. 
This effect is expected to be important in molecular clouds, 
especially in this model as a damping mechanism of Alfv\'en waves
(Kulsrud \& Pearce 1969; Zweibel \& Josafatsson 1983).
It would also change the density structure of a cloud 
lifted up by wave pressure, as demonstrated in the case of 
linear \Alf waves by Martin et al. (1997).

Finally, we believe that a multi-dimensional turbulent 
simulation in a gravitationally stratified medium
is very necessary for the future, especially to obtain 
more definitive dissipation rates in a stratified cloud
and to study the feedback to turbulence due to gravitational
contraction and differential rotation.

\section{Summary}

We have performed a numerical simulation of nonlinear MHD waves in
a stratified molecular cloud. Our main results are as follows:

\begin{enumerate}
\item
Due to the effective pressure of MHD turbulence, our one-dimensional
cloud is lifted upward and establishes a steady-state characterized by
oscillations about a time-averaged equilibrium state. The outer, low
density parts of the cloud undergo the largest oscillations, which
have the character of free-fall motions. After turbulent driving is
discontinued, the cloud falls back toward the initial equilibrium, but
some large-scale oscillations of the cloud are the longest-lived modes.

\item
The nonlinear effect of the wave propagation results in a significant 
conversion of energy from \Alf waves to compressive slow (acoustic) modes 
and consequent shock formation.
However, the energy in transverse modes always remains significantly 
greater than that in longitudinal modes. Further dissipation of 
transverse modes takes place through the generation of smaller scale
structure through nonlinear steepening or a cascade, culminating in 
resistive dissipation on the grid-scale.

\item Within each cloud, the magnetic pressure associated with wave motion,
$\langle B_y^2/(8\pi) \rangle_t$, is in approximate 
equipartition with the kinetic energy in transverse motions 
$\langle \rho (v^\prime_y)^2/2 \rangle_t$ in the region
containing most of the mass. However, in the low density 
outer regions, the wave amplitudes have the
character of standing waves, such that $B_y$ has a node and
$v^\prime_y$ has an antinode. All of this means that the Chandrasekhar-Fermi
model applied to a stratified cloud will have a multiplicative
constant $\alpha$ which is $\approx 1$ with weak spatial dependence through
much of the cloud, but which drops to much lower values near the edge of
the cloud.

\item After turbulent driving is discontinued, the dissipation time
of the cloud turbulence is 
$ \sim 10\, t_0$
or several crossing times
of the time-averaged equilibrium state during turbulent driving. These
times are up to a few times longer than those found in multi-dimensional
periodic simulations. This would be due partly to fewer
dissipation avenues in a one-dimensional approximation, but also 
partly to the generation of long-wavelength modes due to the cloud
stratification. 

\item
For an ensemble of clouds with different levels of internal driving,
we find the relation $\sigma \propto Z^{0.5}$,
where $\sigma$ is the time averaged one-dimensional velocity dispersion 
measured at a size scale $Z$ of the time averaged equilibrium
state of the cloud.

\item
For the same ensemble of clouds, the above relation also implies that
$\sigma \propto \bar{V}_A$,
where $\bar{V}_A$ is the mean Alfv\'en velocity determined from the 
mean density of the cloud. This relation is in agreement with observed
properties of magnetized clouds and cloud cores.

\end{enumerate}

\begin{acknowledgements}

TK acknowledges support due to a Fellowship from the Canadian
Galactic Plane Survey, which is supported by a Natural 
Sciences and Engineering Research Council of Canada (NSERC) Collaborative
Research Opportunities Grant. TK also benefited from a Fellowship from 
SHARCNET, a high-performance computing project funded by ORDCF and 
CFI/OIT. SB was supported by an individual research grant from NSERC.
Numerical computations were carried out mainly on the VPP5000 at 
the Astronomical Data Analysis Center of the National Astronomical 
Observatory, Japan.

\end{acknowledgements}




\begin{deluxetable}{cccccccc} 
\tablecolumns{8}
\tablewidth{0pc}
\tablecaption{
Dissipation time as a function of $\tilde{a}_d$.
Time-averaged height and time-averaged velocity dispersion 
as a function of $\tilde{a}_d$ for two Lagrangian elements:
$z(t=0)=2.51$, approximately the full-mass position, and
$z(t=0)=0.61$, approximately the half-mass position.
}
\tablehead{ 
\colhead{} & \colhead{} & \colhead{} & 
\multicolumn{2}{c}{$z(t=0)=2.51$} &   \colhead{}   & 
\multicolumn{2}{c}{$z(t=0)=0.61$} \\ 
\cline{4-5} \cline{7-8} \\ 
\colhead{$\tilde{a}_d$} & 
\colhead{$t_d/t_0$} & 
\colhead{}  & 
\colhead{$\langle z\rangle_t/H_0$} 
& \colhead{$\langle \sigma^2\rangle_t^{1/2}/c_{s0}$} & 
\colhead{}  & 
\colhead{$\langle z\rangle_t/H_0$} 
& \colhead{$\langle \sigma^2\rangle_t^{1/2}/c_{s0}$}} 
\startdata
10 & 9.0  & & 4.00 & 2.54 &  & 0.76 & 1.18 \\
20 & 5.5  & & 6.82 & 3.73 &  & 1.13 & 1.54 \\
30 & 8.5  & &10.8 & 4.65 &  & 1.78 & 1.92 \\
40 & 9.0  & & 17.9 & 5.73 &  & 2.91 & 2.47 \\ 
50 & 15.0 & & 17.4 & 5.99 &  & 3.34 & 2.68 \\
\enddata
\end{deluxetable} 

\end{document}